\newif\ifonecol

\onecolfalse

  \documentclass[sigconf,screen]{acmart}

  \newcommand{\oneColFigWidth}{\linewidth}
  \newcommand{\doubleColFigWidth}{\linewidth}

\usepackage{amsmath}
\usepackage{stfloats}
\usepackage{multibib}
\usepackage{ enumitem}
\newcites{Appendix}{Appendix References}

  \graphicspath{{pictures/}} 
\AtBeginDocument{%
  \providecommand\BibTeX{{%
    \normalfont B\kern-0.5em{\scshape i\kern-0.25em b}\kern-0.8em\TeX}}}

\copyrightyear{2021}
\acmYear{2021}
\setcopyright{acmlicensed}\acmConference[CHI '21]{CHI Conference on Human Factors in Computing Systems}{May 8--13, 2021}{Yokohama, Japan}
\acmBooktitle{CHI Conference on Human Factors in Computing Systems (CHI '21), May 8--13, 2021, Yokohama, Japan}
\acmPrice{15.00}
\acmDOI{10.1145/3411764.3445356}
\acmISBN{978-1-4503-8096-6/21/05}

\newcommand{\toolName}[1]
{#1}
\newcommand{\system}
{\toolName{Ivy}}
\newcommand{\ivy}
{\system}
\newcommand{\systemlang}
{\toolName{Ivy Template Language}}

\newcommand{\sns}
{\toolName{Sketch-n-Sketch}}
\newcommand{\ivyPolestar}
{IvyPolestar}

\newcommand{\osf}
{\url{https://osf.io/cture/}}

\newcommand{\etal}
{et al.}
\newcommand{\etals}
{et al.'s}

\newcommand{\ie}{{i.e.}}
\newcommand{\eg}{{e.g.}}

\newcommand{\cf}{{cf.}}

\ifx\hideComments\undefined
  \newcommand\rc[1]{{\color{blue}[RC: #1]}}
  \newcommand\am[1]{{\color{red}[AM: #1]}}
  \newcommand\glk[1]{{\color{orange}[GLK: #1]}}
  
\else
  \newcommand\rc[1]{{}}
  \newcommand\am[1]{{}}
  \newcommand\glk[1]{{}}
  
\fi

\newcommand{\authorVersion}{HIDE EM} 
\ifx\authorVersion\undefined
  \newcommand\spaceit[0]{\vspace{-0.2in}}
\else
  \newcommand\spaceit[0]{{}}
\fi

\newcommand{\secref}[1]{\hyperref[#1]{Sec.~\ref*{#1}}}
\newcommand{\appendixref}[1]{\hyperref[#1]{Appendix.~\ref*{#1}}}
\newcommand{\figref}[1]{\hyperref[#1]{Fig.~\ref*{#1}}}
\newcommand{\eqnref}[1]{\hyperref[#1]{Eqn.~\ref*{#1}}}
\newcommand{\tabref}[1]{\hyperref[#1]{Table ~\ref*{#1}}}

{\par\kern\topsep}

\newcommand{\footnoteWithIndent}[1]
{\footnote{#1}}

\def\subsubsec#1
{\subsubsection{#1}}

\newcommand{\jField}[1]
{\texttt{"#1"}}
\newcommand{\jString}[1]
{\texttt{"#1"}}

\newcommand{\jParam}[1]
{\texttt{#1}}

\newcommand{\goal}[1]
{\textbf{G#1}}

\begin{document}

\title
[Integrated Visualization Editing via Parameterized Declarative Templates]
{Integrated Visualization Editing via
  \ifonecol
  \else
    \\
  \fi
  Parameterized Declarative Templates}

\author{Andrew McNutt}
\orcid{0000-0001-8255-4258}
\affiliation{%
  \institution{University of Chicago}
  \city{Chicago}
  \state{IL}
  \postcode{60615}
}
\author{Ravi Chugh}
\affiliation{%
  \institution{University of Chicago}
  \city{Chicago}
  \state{IL}
  \postcode{60615}
}

\begin{abstract}%
    Interfaces for creating visualizations typically embrace one of several common forms.
    \emph{Textual specification} enables fine-grained control,
    \emph{shelf building} facilitates rapid exploration, while
    \emph{chart choosing} promotes immediacy and simplicity.
    Ideally these approaches could be unified to integrate the user- and usage-dependent benefits found in each modality, yet these forms remain distinct.

    We propose \emph{parameterized declarative templates}, a simple abstraction mechanism over JSON-based visualization grammars, as a foundation for multimodal visualization editors.
    We demonstrate how templates can facilitate organization and reuse by factoring the more than 160 charts that constitute Vega-Lite's example gallery into approximately 40 templates.
    We exemplify the pliability of abstracting over charting grammars by implementing---as a template---the functionality of the shelf builder Polestar (a simulacra of Tableau)
    and a set of templates that emulate the Google Sheets chart chooser.
    We show how templates support multimodal visualization editing by implementing a prototype and evaluating it through an approachability study.
\end{abstract}

\ccsdesc[500]{Human-centered computing~Visualization systems and tools}
\ccsdesc[500]{Human-centered computing~Graphical user interfaces}

\keywords{%
  Information Visualization,
  Declarative Grammars, Templates,
  User Interfaces,
  Ivy, Systems
}

\maketitle

\begin{figure*}[t]
    \centering
    \includegraphics[width=\linewidth]{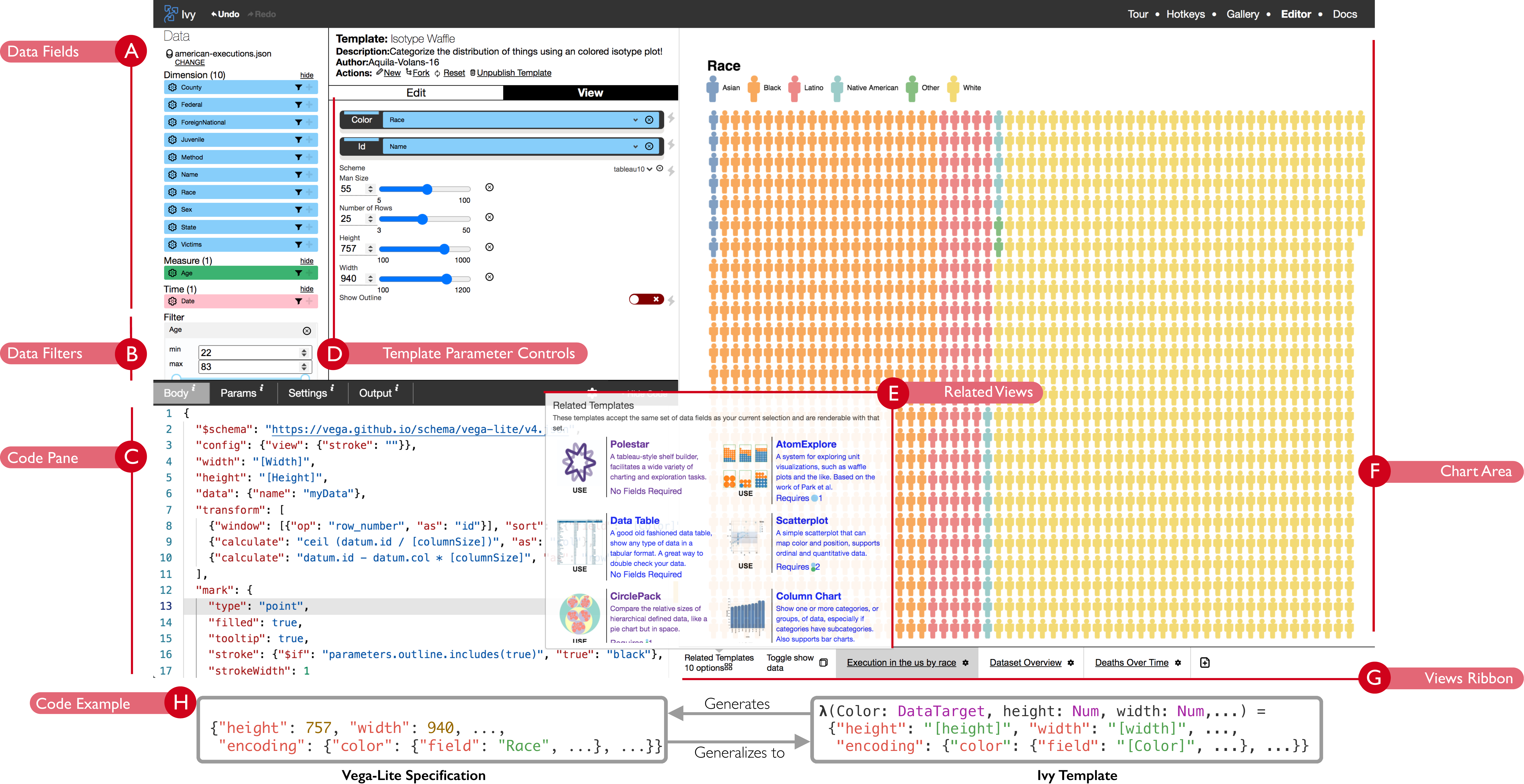}
    \caption{
        The \system{} visualization editor combines textual specification with GUI-based shelf building and chart choosing.
        Here, an ``Isotype Waffle'' \emph{template} is used  visualize the people executed in the USA under the death penalty by race since 1977 \cite{chandler_executions_2018}.
        Templates are functions with typed parameters that abstract JSON
        specifications in declarative visualization grammars.
    }
    \Description[An annotated view of our ivy web application]{The screen is vertically divided into two sections, on the right an isotype of people executed by the death penalty colored by race is shown. The left side is horizontally divided into two sections, the top showing two columns one containing various data fields from the dataset, and the right showing controls for configuring the chart on the right. The bottom of the vertical split shows a code editor showing the JSON creating the chart. There is a pop-up overlaid on top of the middle of the chart showing the related templates popover. Below the screenshot are two pieces of code, the left showing Vega-Lite code and the right showing an ivy template that generated that code. There are arrows indicating that the ivy template generates the Vega-Lite specification, while the Vega-Lite specification generalizes to the Ivy template.}
    \label{fig:annotated-interface}
    \spaceit{}
\end{figure*}

\section{Introduction}
\label{sec:intro}

Every user interface design involves compromise. Which tasks should be made easy at the expense of making other tasks cumbersome or even impossible?

There are several common user interface modalities for creating visualizations, each with distinct trade-offs \cite{grammel_survey_2013}.
\textbf{\emph{Chart choosers}} (as in Excel) allow users to rapidly construct familiar visualizations at the expense of flexibility.
\textbf{\emph{Shelf builders}} (as in Tableau) facilitate dynamic exploration but can obstruct the construction of specific chart forms or the addition of visual nuances.
\textbf{\emph{Textual programming}}
is highly expressive but can impede rapid exploration.

\begin{figure}[t]
    \begin{minipage}{\oneColFigWidth}
        \centering
        \includegraphics[width=\linewidth]{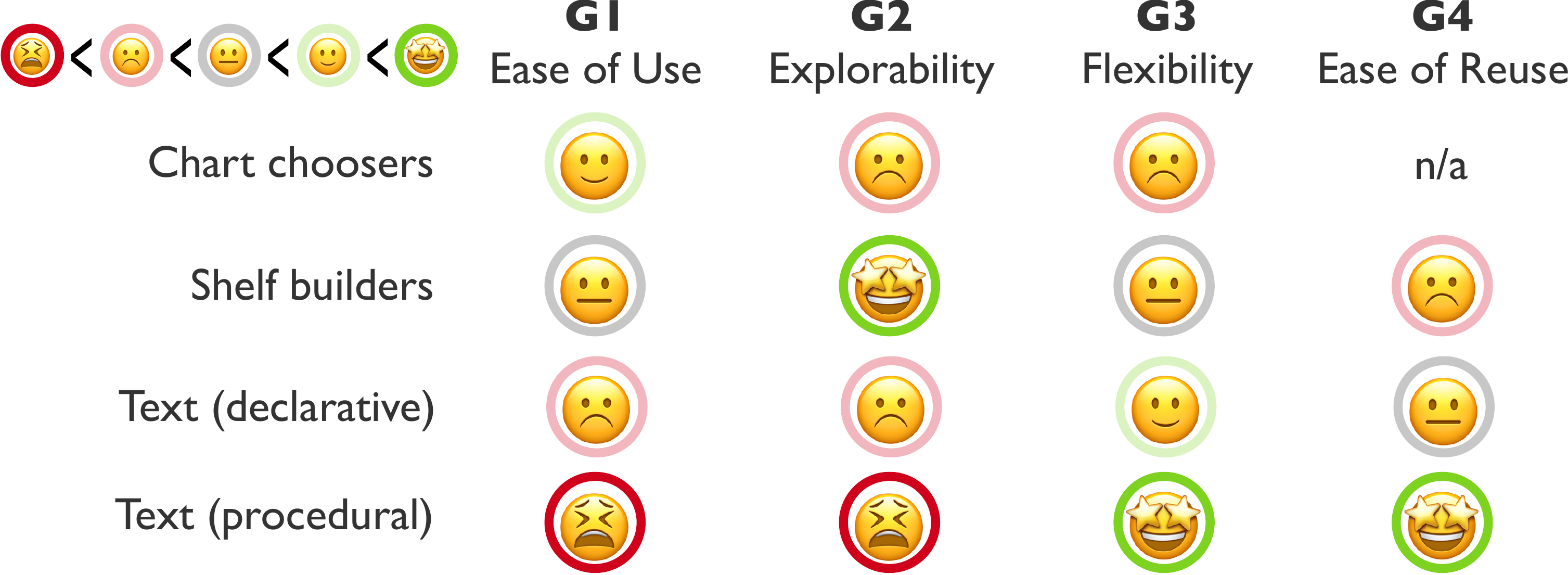}
        \caption{
            Chart making modalities have distinct strengths and weaknesses.
            Using \emph{parameterized declarative templates}, \ivy{} strives to combine the strengths of several common modalities in this design space.
        }
        \Description[A matrix of smiley face emojis]{A matrix of smiley face emojis highlighting the relative performance of a variety of chart making modalities on a number of tasks.}
        \label{fig:pro-cons-cartoon}
    \end{minipage}
    \spaceit{}
\end{figure}

A modality which may be well suited to the initial stages of a task, may become cumbersome in subsequent phases.
For example, while a low-configuration approach for rapid data exploration might suffice at the beginning of their work, the user might subsequently require a more flexible---even if higher friction---interface that allows them to tune specific details of their chart.
Graphical user interface (GUI) systems, such as chart choosers and shelf builders, can leave experts wanting more precise control over the chart creation process, while textual systems can leave novices in need of assistance.
Without the ability to move between modalities, users are stuck with each interface's shortcomings.
None of the existing single-mode interfaces simultaneously achieve each of several goals (as in \figref{fig:pro-cons-cartoon}): ease of use~(\goal{1}), explorability~(\goal{2}), flexibility~(\goal{3}), and ease of reuse~(\goal{4}).
These deficits can force users to switch tools across their analysis~\cite{hografer_revize_2019},
or compel proficient users to seek ad hoc solutions that are difficult to repeat in different contexts.

Ideally, interfaces of varying complexity could be integrated such that both novice users (for whom chart choosers are often best suited \cite{grammel_how_2010}) and experts (whose most profitable interface will vary) obtain the benefits of each modality as their tasks require.
Unfortunately this territory remains under-explored, as visualization systems tend to prefer one-size-fits-all designs.
Declarative visualization grammars are an enticing starting point as they provide significant flexibility for specifying visualizations as text~(\goal{3}). However, they lack the abstraction mechanisms found in full-featured programming languages.
This paper considers the question:

\textit{%
    Can we extend declarative grammars with abstraction mechanisms for reuse (\goal{4}), in a way that facilitates explorability (\goal{2}) as in shelf builders and ease of use (\goal{1}) as in chart choosers?
}%

To answer to this question, we propose an abstraction mechanism called \textbf{\textit{parameterized declarative templates}}, or simply \textbf{\textit{templates}},
which extend textual, declarative grammars---specifically, those in which specifications are defined using JavaScript Object Notation~(JSON)~\cite{ecma_json_2017}---with mechanisms for reusing chart definitions.
As depicted in \figref{fig:annotated-interface}h, templates abstract ``raw'' declarative specifications with parameters that specify data fields for a visual encoding (\eg{}~\jParam{Color}) and design parameters (\eg{}~\jParam{height} and \jParam{width}), making them more easily reused (\goal{4}).
Templates can be rapidly instantiated for different data sets, shared among communities, and modified to taste---alleviating barriers to opportunistically~\cite{brandt_two_2009} leveraging the rich body of grammar-based charts found online.
Templates, which are essentially functions in a simple programming language with variables and conditionals, are described formally in \secref{sec:lang-design}.

We systematically apply this idea in a prototype visualization editor, called \ivy{}, in which templates are created and instantiated through text- and GUI-based manipulation (\goal{1}).
Templates abstract over arbitrary JSON-based visualization grammars---%
our implementation currently supports Vega, Vega-Lite, Atom~\cite{park_atom_2018}, and a toy data table language---%
allowing users to easily move between them (\goal{3}).
We implemented an \ivy{} template that recreates the functionality of the shelf builder system Polestar~\cite{wongsuphasawat_voyager_2015, wongsuphasawat_voyager_2017}, that serves as the default view of the system, facilitating explorability (\goal{2}).
Our user interface design is described in \secref{sec:ui}.

The systematic application of templates furthermore enables two notable interaction features beneficial to exploration (\goal{2}).
\emph{Catalog search} utilizes standard type-based compatibility checks to implement an extensible---if simple---recommendation system.
\emph{Fan out} facilitates rapid exploration within a template by juxtaposing multiple chart configurations on demand.
These features are described in \secref{sec:emergent}.

To evaluate how templates may serve as a foundation for multimodal visualization editing, we performed two investigations.
We considered how they might usefully reproduce and compress extant families of charts, first by factoring the 166 unique examples that constitute the Vega-Lite gallery into 43 templates, and then by reconstructing the 32 charts of the Google Sheets chart chooser as 16 templates.
We then conducted a small approachability study of our prototype \ivy{} system which demonstrated that, with some training and guidance, users are able to create and instantiate templates by mixing the available modalities.
We present these results in \secref{sec:eval}.

Although the main ideas regarding templates---abstraction over value definitions---and their instantiations
are familiar concepts, the contribution of this paper is to explore and evaluate how these ideas may help integrate what are currently disparate approaches for creating visualizations.
The result of this exploration is our Ivy prototype, which exhibits the promise of this approach.
This composition of modalities provides
several potential opportunities,
including borrowing and adapting external examples (which many shelf builders lack),
rapid exploration of parameter combinations (which programmatic interfaces often lack)
and self-service chart creation  (which chart choosers typically lack).
Compared to textual programming in existing declarative grammars, the abstraction layer provided by templates
simplifies the process of chart reuse and therein reduces code clones (as in our chart gallery reproductions).

Our prototype is available at \url{https://ivy-vis.netlify.app/}.
The appendix further details our evaluation and implementation, while other supplementary materials can be found at \osf{}.

\section{Related Work}
\label{sec:related}

Software systems for creating visualizations can be classified into a variety of interaction modalities~\cite{grammel_survey_2013, mei_design_2018}.
Among these, we aim to bridge three of the most common: chart choosing, shelf building, and textual programming.
We now discuss declarative grammars and visualization editors, as they relate to the key ideas in this paper:
(1)~to endow declarative visualization grammars with basic abstraction mechanisms, and
(2)~to design a multimodal UI based on templates for creating and editing visualizations.

\subsection{Declarative Visualization Grammars}

Declarative grammars of graphics have proven extremely popular \cite{mackinlay_automating_1986, wilkinson_grammar_2013, wickham_layered_2010, stolte_polaris_2002, hanrahan_vizql_2006, wang_visualization_2019, vanderplas_altair_2018,  pu_probabilistic_2020, satyanarayan_vega-lite_2016, guozheng_gotree_2020, satyanarayan_declarative_2014, satyanarayan_reactive_2016, tao2020kyrix, wickham_product_2011, park_atom_2018,kim2021Gemini,li2021P6, poorthuis2020florence, lavikkagenomespy},
with different approaches providing more expressiveness than others for particular tasks.
Compared to full-featured, procedural visualization programming languages \cite{hunter_matplotlib_2007,bostock_protovis_2009,bostock_d3_2011},
declarative visualization languages trade fine-grained control over how to render a view for concise, expressive means to specify what to render.
The idea behind parameterized declarative templates is to tilt these specification languages slightly back towards the full-featured languages, by adding some basic programming abstraction mechanisms.

Many declarative grammars~\cite{satyanarayan_declarative_2014, satyanarayan_reactive_2016, satyanarayan_vega-lite_2016, park_atom_2018, tao2020kyrix, lavikkagenomespy, kim2021Gemini,li2021P6} adopt JSON as their specification format, which allows chart definitions to be easily consumed by various environments and tools.
Vega and its surrounding ecosystem exemplify these benefits, which we seek to amplify by making JSON-based grammar specifications easier to reuse and explore.
A related effort is Harper \etals~\cite{harper_converting_2018} system for converting D3 charts \cite{bostock_d3_2011} into reusable Vega-Lite specifications, which they also dub templates.
Their templates provide abstractions related to ours, but are restricted to a limited subset of D3 charts.
\ivy{} templates abstract arbitrary JSON and thus support any JSON-based visualization
grammar.

\subsection{User Interfaces for Visualization}

\subsubsec{Chart Choosers}
This prevalent technique starts with selecting a desired chart form from a (potentially large) set of chart types, and then customizing it through a (usually limited) set of options.
They are ubiquitous among analytics tools, such as spreadsheets,  and are often found in visual analysis (VA) environments~\cite{datawrapper_datawrapper_nodate, nteract_data-explorer_2020, tibco_spotfire_2020, ahlberg_spotfire_1996}.
This workflow facilitates an approach to charting held by many novices~\cite{grammel_how_2010},
but can also lead to premature commitment~\cite{green1989cognitive}.

\ivy{} provides a gallery of templates, some of which constitute single distinct chart forms, as in conventional chart choosers.
Some systems, like \ivy{}, provide APIs for extending the selection of charts \cite{mauri_rawgraphs_2017, noauthor_flourish_nodate}, but, unlike \ivy{}, they tend not to allow these changes from within the tool.
Some systems \cite{viegas_many_2007, datawrapper_datawrapper_nodate, mauri_rawgraphs_2017} provide social features, wherein users can create, share, and modify charts.
\ivy{} users can publish, fork, and remix templates through a shared template server.

\subsubsec{Shelf Builders}
In shelf builders, users map data columns to visual attributes, typically in a manner that is motivated by a principled visualization framework, such as VizQL~\cite{hanrahan_vizql_2006} or the \emph{Grammar of Graphics}~\cite{wilkinson_grammar_2013}.
Tableau~\cite{stolte_polaris_2002} and Charticulator~\cite{ren2018charticulator} are prominent examples of this paradigm.
Data fields of the current dataset are often represented as draggable ``pills" that can be placed onto ``shelves,'' each of which denotes different aspects of the visualization (such as horizontal position or color).

We designed the GUI controls in \ivy{} for choosing arguments for template parameters to resemble the visual conventions of Polestar~\cite{wongsuphasawat_voyager_2015, wongsuphasawat_voyager_2017}, itself a facsimile of Tableau~\cite{stolte_polaris_2002}.
Shelf builders support a range of tasks, including presentation~\cite{carr_lyra_2014, saket_visualization_2016, hoffswell_techniques_2020, hu_dive_2018} and exploration~\cite{stolte_polaris_2002, wongsuphasawat_voyager_2015, wongsuphasawat_voyager_2017}.
We facilitate the latter by constructing an \ivy{} template---\ivyPolestar{}---that emulates and expands upon the Polestar application.
This ``default'' template provides a familiar shelf builder interface upon application startup.

Many VA tools are backed by JSON-based declarative grammars (such as Lyra~\cite{carr_lyra_2014, zong2021lyra} or Voyager~\cite{wongsuphasawat_voyager_2017,wongsuphasawat_voyager_2015}), yet do not allow users to modify the underlying specifications.
Restricting modifications to what can be created within such tools misses a significant opportunity, as there is a wealth of online knowledge and examples that ought to be utilized \cite{brandt_two_2009}.
In commercial VA environments (such as Tableau), the disconnect from the underlying specification can cause fine-grained editing capabilities to be relegated to deeply-nested drop-down menus, which can preclude feature utilization.
We address these issues by making our templates malleable, so that users can adapt the interface to their needs.

\subsubsec{View Exploration in Shelf Builders}
\label{sec:recommendation}

Shelf builders often prominently feature affordances for rapid data exploration, often in the form of recommendation systems~\cite{wongsuphasawat_voyager_2015,wongsuphasawat_voyager_2017}. Lee describes the current state of the art of these subsystems~\cite{lee_insight_2020}.
Our template-based architecture gives rise to simplified analogs to these ``smart'' features, but which still facilitate view exploration goals.

Our \emph{catalog search} is an extensible variation of Tableau's Show Me feature~\cite{mackinlay_show_2007}, enabling users to create examples that are available for subsequent recommendation.
Like Show Me, catalog search uses data-role based matching to identify potential alternatives.
Unlike Show Me, however, catalog search does not internalize domain-specific knowledge of the underlying grammars---the recommendations do not consider, for example, known relationships between human perception and chart configuration.

Our \emph{fan out} enables users to rapidly search across both data and design space by simultaneously juxtaposing parameter selections of interest.
Similarly to catalog search, this feature is simple but achieves many of the same goals as additional prior systems~\cite{noauthor_morph_nodate, agarwal2019viswall, ma2000visualizing} that often include sophisticated domain-specific knowledge, by enabling users to rapidly explore alternatives in a low cost manner (\ie{}~in a simple design gallery~\citep{marks1997design}) and to manipulate collections of examples simultaneously (\`a la Juxtapose \cite{hartmann2008design}).
This technique most closely resembles Ren\'e's~\cite{reneJonGoldBlogPost} combinatorial design exploration, although extended to encompass data-based variations.

\subsubsec{Text-Based Editors}
There is a long history of UIs that combine textual chart programming with a system for rendering those charts.
Computational notebooks, such as Jupyter Notebook~\cite{kluyver_jupyter_2016}, facilitate chart construction through tight feedback loops between code and rendered charts.
In a similar manner to \ivy{} templates, papermill~\cite{nteract_papermill_2020} allows analysts to parameterize computational notebooks which can then be run in a non-notebook environment.
\toolName{Observable}~\cite{bostock_better_2018} provides a reactive-programming platform for creating rich web-based visualizations, which users are encouraged to fork and remix.
Wood \etal~\cite{wood_design_2019} take a literate programming approach to the visualization design process by enabling authors to build Vega and Vega-Lite charts in a Markdown document.

Most closely related to our work is the \toolName{Vega-IDE}~\cite{noauthor_editoride_nodate}, which augments textual specification of \toolName{Vega} and \toolName{Vega-Lite} charts with debugging tools \cite{hoffswell_augmenting_2018}, and Chart Builder~\cite{dataworld_chart_nodate}, which allows users to edit Vega-Lite charts through a GUI or text, but not both.
We borrow much of \toolName{Vega-IDE} interface design in the text-editing portion of our system;
for instance, we make use of Microsoft Monaco's JSON-Schema \cite{pezoa_foundations_2016} based support for JSON grammars as a way to provide linting and validation.
Beyond standard text-editing features, we also implement extensible heuristic rewrite rules that (automatically) suggest abstractions of JSON values into parameters.
This eases the flow of converting a declarative specification into a reusable template.

\subsubsec{Multimodal Editors}
Several prior works have explored combinations of user interface modalities for creating visualizations.
\toolName{Liger}~\cite{saket_liger_2019} mixes together shelf-based chart specification and \emph{visualization by demonstration}.
\toolName{Hanpuku}~\cite{bigelow_iterating_2016}, Data-Driven Guides~\cite{kim_data-driven_2016}, and \toolName{Data Illustrator}~\cite{liu_data_2018} combine visual editor-style manipulation with visual chart specification or textual programming.
Victor~\cite{victor_drawing_2013} explored a prototype that combined a spreadsheet with direct manipulation and manual view specification. His system enabled highly expressive visualization construction, whereas we focus on supporting analytic tasks.

Tools such as Jupyter Widgets~\cite{jupyterWidgets} and Shiny~\cite{shiny} provide mechanisms for parameterizing analysis code in an ad hoc manner, however these lack many of the graphical affordances for exploration found in many visual analytics systems.

Systems including \toolName{mage}~\cite{kery2020mage}, \toolName{Wrex}~\cite{drosos2020Wrex}, and \toolName{B2} \cite{wu2020B2} expand on these ideas by intermingling text and graphical specification in computational notebooks.
\sns{}~\cite{hempel_sketch-n-sketch_2019} takes a bidirectional approach to combining (non data-driven) visual editing with textual programming (in a full-featured, procedural language).
ReVize~\cite{hografer_revize_2019} seeks to support multiple modalities by chaining together analysis tools though a Vega-Lite-based API.

Our investigation in this paper---to integrate chart choosing, shelf building, and textual specification---is complementary to these efforts.
This combination of modalities offers a rich feature-space (such as our \emph{catalog search} and \emph{fan out} features), enables educational opportunities (by presenting a synchronized view between GUI and perhaps unfamiliar textual grammars), and supports a variety of data exploration tasks.

\section{Motivating Use Cases}
\label{sec:axel-tabitha}

Here we describe how two hypothetical users---Axel, a novice user of visual analytics systems, and Tabitha, a visual analytics power user---carry out two example workflows in \ivy{}.
A motivation for our work is to allow users like Axel and Tabitha to collaborate and benefit from each other's efforts by working within a common system---%
aiming to supplant single-modality tooling that contributes to \emph{designer-developer breakdowns}~\citep{Enact}.

\subsection{Axel: A Novice User}\label{sec:novice}
Axel wants to make a chart showing death penalty executions in the USA for a report he is writing.
After loading a relevant dataset, Axel browses templates that have been created by others.
Axel benefits from how other members of his team can easily grow the collection of templates to adapt to the team's changing visualization needs.

Unsure of what values are in the dataset, he selects a univariate bar chart template and uses fan out (\secref{sec:ui-fan-out}) to view all of the dimension fields simultaneously. As he views these summaries, he gradually removes fan out options that are not interesting or do not serve his task. He eventually selects the option to display \jString{race}. He notices that there are 16 related templates (\secref{sec:ui-catalog-search}) that he could use. The ``Isotype Waffle'' catches his eye. He selects it
to view a fully formed Isotype colored by \jString{race}.
The chart does not have quite the right dimensions so he adds \jParam{width} and \jParam{height} variables to the template, which allows him to make small incremental adjustments from the GUI to produce the final version (\figref{fig:annotated-interface}f). He takes a screenshot and adds it to his report.

\subsection{Tabitha: An Expert User}\label{sec:expert}

Tabitha is interested in exploring the Gapminder dataset \cite{rosling_health_2011}, so she loads the data and selects the \ivyPolestar{} template. She explores the data using the familiar row and column abstractions (as in \figref{fig:polestar}) to investigate iterative hypotheses.
After viewing several scatter plots and bar charts, Tabitha wants to see a hierarchical representation of the part-to-whole relationships between region, country, and GDP; a Sunburst chart comes to mind.

Unable to find a satisfactory chart in the \ivy{} gallery, she browses the Vega gallery and finds one, but the way that it works with data does not quite match her intended usage.
She creates a blank template in \ivy{} and pastes in the example from the Vega gallery. She is then presented with a series of automatically generated suggestions (\secref{sec:text-editing}) on how she might templatize her chart---clicking through these creates new shelves as appropriate. She text-edits the data transformation logic to accommodate her desired functionality.
She uses the built-in debugging tools to view the results of the current data state, iteratively developing her transformations.
She adds parameters for controlling width, height, color scheme, and other aesthetic values, at which point she is happy with the template and decides to share it.
She clicks the publish button to make the template available on the community server, ensuring that she and others can reuse her work in the future.
Finally, she instantiates the Sunburst for her dataset and takes a screenshot.

\section{System Design}
\label{sec:system}

\begin{figure*}[tb]
    \centering
    \includegraphics[width=\doubleColFigWidth]{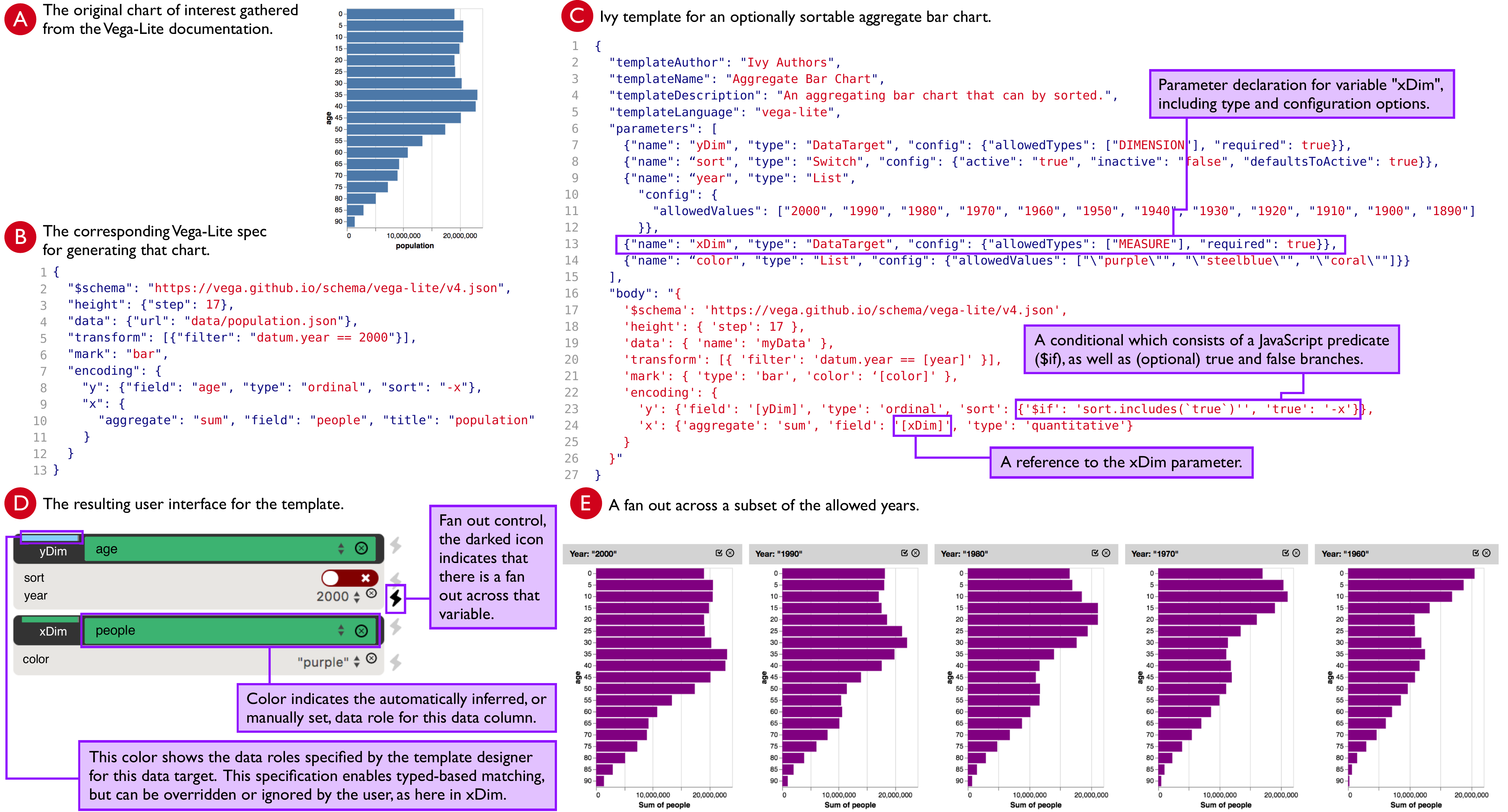}
    \caption{
        Users can transform example chart specifications into reusable templates.
        Here we show the process of taking
        (A)~a bar chart of interest and
        (B)~its corresponding Vega-Lite spec, and transforming it into
        (C)~an \system{} template. Using
        (D)~the corresponding GUI shelf builder pane, the user can
        (E)~fan out across parameters of interest to see alternatives simultaneously.
    }
    \Description[An example of how the ivy language works illustrated by several screenshots.]{An example of how the ivy language works illustrated by several screenshots. A) shows a vertical blue bar chart of population (x axis) vs age (y axis), and is labeled "the original chart of interest gathered from the Vega-Lite documentation". B) shows the Vega-Lite code that created the chart in part A, and is labeled "The corresponding Vega-Lite spec for generating that chart." C) Shows An Ivy template that can create the Vega-Lite code from part B) it is labeled "Ivy template for an optionally sortable aggregate bar chart. " D) Shows a GUI created by using the Ivy Template in part C it is labeled "The resulting user interface for the template." E) Shows a series of variations on the bar chart from part A, although it is now purple and consists of only data from 5 different decades.}
    \label{fig:bar-chart}
\end{figure*}

\newcommand{\sep}{\hspace{0.06in}}
\newcommand{\miniSepOne}{\hspace{0.01in}}

\newcommand{\figSyntaxLineBreak}{\\[1pt]}
\newcommand{\figSyntaxSpaceNextCategory}{\\[1pt]}
\newcommand{\figSyntaxSpaceNextCategoryMoreSpace}
{\\[6pt]}
\newcommand{\figSyntaxSpaceItem}{\sep\mid\sep}
\newcommand{\figSyntaxSpaceItemNarrow}{\mid}
\newcommand{\figSyntaxEnd}
{\end{array}$}

\newcommand{\figSyntaxBegin}
{$\begin{array}{rrcll}}
\newcommand{\figSyntaxRowLabel}[2]
{{\textbf{#1}}\!&\!{#2}&{::=}&}
\newcommand{\figSyntaxRow}[1]
{\textit{#1}&&\miniSepOne\mid\miniSepOne&}

\newcommand{\vsepRuleHeight}
{0.05in}
\newcommand{\vsepRule}{\vspace{\vsepRuleHeight}}

\newcommand{\relDescription}[1]{\ensuremath{\textrm{\textbf{#1}}}}
\newcommand{\JudgementBox}[1]{{
            \setlength{\fboxsep}{3pt} 
            \fbox{#1}
        }}
\newcommand{\judgementHead}[2]
{\ensuremath{\relDescription{#1}\hfill\JudgementBox{$#2$}}}
\newcommand{\judgementHeadTwo}[3]
{\ensuremath{\relDescription{#1}\ \textrm{#2}\hfill\JudgementBox{$#3$}}}
\newcommand{\judgementHeadNameOnly}[1]
{\ensuremath{\relDescription{#1}\hfill}}

\newcommand{\varTemplate}{t}
\newcommand{\varAtom}{a}
\newcommand{\varStr}{s}
\newcommand{\varNum}{n}
\newcommand{\varBool}{b}
\newcommand{\varCond}{p}
\newcommand{\varExp}{e}
\newcommand{\varVal}
{j}
\newcommand{\varVar}{x}
\newcommand{\varSymbol}
{y}
\newcommand{\varType}{T}
\newcommand{\varRole}{R}
\newcommand{\varConfig}
{\textsl{args}}
\newcommand{\varSettings}{S}

\newcommand\mydots{\hbox to 1em{.\hss.\hss.}}
\newcommand{\varSpecLang}
{\mathcal{L}}
\newcommand{\varView}
{v}

\newcommand{\ttlcurly}{\ensuremath{\texttt{\char`\{}}\hspace{0.00in}}
\newcommand{\ttrcurly}{\ensuremath{\hspace{0.00in}\texttt{\char`\}}}}
\newcommand{\ttlparen}{\ensuremath{\texttt{(}}}
\newcommand{\ttrparen}{\ensuremath{\texttt{)}}}
\newcommand{\ttlbrack}{\ensuremath{\texttt{[}}}
\newcommand{\ttrbrack}{\ensuremath{\texttt{]}}}
\newcommand{\ttpipe}{\ensuremath{\texttt{|}}}
\newcommand{\ttdash}{\ensuremath{\texttt{-}}}


\newcommand{\templateComponent}[1]
{\textsl{#1}}
\newcommand{\templateComponentDecl}[2]
{\templateComponent{#1} = #2}
\newcommand{\templateComponentProj}[2]
{#1.\templateComponent{#2}}
\newcommand{\templateFunction}[2]
{\lambda{#1}\texttt{.} \hspace{0.02in} {#2}}
\newcommand{\varDecls}[2]
{\texttt{(}#1\texttt{:}#2\texttt{,} \mydots\texttt{)}}
\newcommand{\varDeclsN}[4]
{\texttt{(}#1\texttt{:}#2\texttt{,} \mydots \texttt{,} \hspace{0.02in} #3\texttt{:}#4\texttt{)}}
\newcommand{\symbolDecls}[1]
{\texttt{[}#1\texttt{,} \mydots\texttt{]}}
\newcommand{\symbolDeclsN}[2]
{\texttt{[}#1\texttt{,} \mydots \texttt{,} \hspace{0.02in} #2{]}}
\newcommand{\jRecordBind}[2]
{#1 \texttt{:} #2}
\newcommand{\jRecordOne}[2]
{\ttlcurly \hspace{0.02in} \jRecordBind{#1}{#2} \hspace{0.02in} \ttrcurly}
\newcommand{\jRecord}[2]
{\ttlcurly \hspace{0.02in} \jRecordBind{#1}{#2} \texttt{,} \mydots \hspace{0.00in} \ttrcurly}
\newcommand{\jRecordN}[4]
{\ttlcurly \hspace{0.02in} \jRecordBind{#1}{#2} \texttt{,} \mydots \texttt{,}
    \hspace{0.00in} \jRecordBind{#3}{#4} \hspace{0.0in} \ttrcurly}
\newcommand{\jRecordBindSingle}[2]
{(#1 \texttt{:} #2)}
\newcommand{\jRecordCat}[2]
{#1 \hspace{0.02in} \cup \hspace{0.02in} #2}
\newcommand{\jRecordCatThree}[3]
{\jRecordCat{#1}{\jRecordCat{#2}{#3}}}
\newcommand{\jList}[1]
{\texttt{[} \hspace{0.02in} #1\texttt{,} \mydots \hspace{0.02in} \texttt{]}}
\newcommand{\jListN}[2]
{\texttt{[} \hspace{0.02in} #1\texttt{,} \mydots \texttt{,}
\hspace{0.02in} #2 \hspace{0.02in} \texttt{]}}
\newcommand{\ite}[3]
{\texttt{if}\ #1\ (\texttt{then}\ #2)\ \texttt{else}\ #3}
\newcommand{\iteOptional}[3]
{\texttt{if}\ #1\ \texttt{then}\ #2\ (\texttt{else}\ #3)}
\newcommand{\iteNoElse}[2]
{\texttt{if}\ #1\ \texttt{then}\ #2}

\newcommand{\tTag}[1]
{\texttt{#1}}
\newcommand{\tTagAndConfig}[1]
{\texttt{#1}\ \varConfig}
\newcommand{\tDataTarget}
{\tTagAndConfig{DataTarget}}
\newcommand{\tMultiDataTarget}
{\tTagAndConfig{MultiDataTarget}}

\newcommand{\settings}[2]
{#1 \mapsto #2, \hspace{0.02in} \mydots}

\newcommand{\settingsN}[4]
{#1 \mapsto #2, \hspace{0.02in} \mydots, \hspace{0.02in} #3 \mapsto #4}

\definecolor{measureColor}{HTML}{3CB371}
\definecolor{dimensionColor}{HTML}{87CEFA}
\definecolor{timeColor}{HTML}{FFC0CB}

\newcommand{\role}[1]
{\texttt{#1}}
\newcommand{\roleMeasure}{
    \role{Measure}
    \hspace{0.05in}
    {\color{measureColor}{
            \text{\raisebox{3pt}{\circle*{6}}}
        }}
    \hspace{-0.07in}
}
\newcommand{\roleDimension}{
    \role{Dimension}
    \hspace{0.05in}
    {\color{dimensionColor}{
            \text{\raisebox{3pt}{\circle*{6}}}
        }}
    \hspace{-0.07in}
}
\newcommand{\roleTime}{
    \role{Time}
    \hspace{0.05in}
    {\color{timeColor}{
            \text{\raisebox{3pt}{\circle*{6}}}
        }}
    \hspace{-0.07in}
}

\newcommand{\specLang}[1]
{\texttt{#1}}
\newcommand{\specLangVega}{\specLang{Vega}}
\newcommand{\specLangVegaLite}{\specLang{Vega-Lite}}
\newcommand{\specLangAtom}{\specLang{Atom}}
\newcommand{\specLangTable}{\specLang{Table}}

\newcommand{\applyTemplate}[2]
{\ensuremath{#1}({#2})}
\newcommand{\applySubst}[2]
{\ensuremath{#1}{#2}}
\newcommand{\evalLang}[2]
{\ensuremath{\llbracket{#2}\rrbracket_{\hspace{0.01in}\mathsf{#1}}}}
\newcommand{\evalIvy}[1]{\evalLang{\ivy}{#1}}
\newcommand{\evalJS}[1]{\evalLang{JS}{#1}}


\begin{figure*}[t]
    \ifonecol
        \hspace{-0.1in}
    \fi
    \centering
    \begin{minipage}{
            0.5\textwidth
        }
        \ifonecol 
            \small
        \fi
        %
        %
        %
        \figSyntaxBegin
        \figSyntaxRowLabel{Templates}{\varTemplate}
        \hspace{-0.05in} 
        \Bigg\{
        \begin{array}{l}
            \templateComponentDecl{func} {\templateFunction{\varDecls{\varVar_1}{\varType_1}}{\varExp}} \\
            \templateComponentDecl{lang}{\varSpecLang}, \templateComponentDecl{metadata}{\cdots},       \\
            \templateComponentDecl{symbols}{\symbolDecls{\varSymbol_1}},\

            \\
        \end{array}
        %
        \\ \\[-5pt] 
        \figSyntaxRowLabel{Atomic Values}{\varAtom}
        \varStr
        \figSyntaxSpaceItem
        \varNum
        \figSyntaxSpaceItem
        \varBool
        \figSyntaxSpaceNextCategory
        \figSyntaxRowLabel{JSON Values}{\varVal}
        \varAtom
        \figSyntaxSpaceItem
        \jRecord{\varStr_1}{\varVal_1}
        \figSyntaxSpaceItem
        \jList{\varVal_1}
        \figSyntaxSpaceNextCategory
        \figSyntaxRowLabel{Expressions}{\varExp}
        \varAtom
        \figSyntaxSpaceItem
        \jRecord{\varStr_1}{\varExp_1}
        \figSyntaxSpaceItem
        \jList{\varExp_1}
        \figSyntaxLineBreak
        \figSyntaxRow{Variables}
        \varVar
        \figSyntaxSpaceItem
        \varSymbol
        \figSyntaxLineBreak
        \figSyntaxRow{Conditionals}
        \iteOptional{\varCond}{\varExp_1}{\varExp_2}
        \figSyntaxLineBreak
        \figSyntaxRowLabel{Predicates}{\varCond}
        \varStr
        \quad (\textrm{where}\ \evalJS{\varStr} = \varBool)
        %
        \figSyntaxSpaceNextCategoryMoreSpace
        \figSyntaxRowLabel{Settings}{\varSettings}
        \settings{\varVar_1}{\varAtom_1}
        %
        \figSyntaxSpaceNextCategoryMoreSpace
        \figSyntaxRowLabel{Data Roles}{\varRole}
        \roleMeasure
        \figSyntaxLineBreak
        \figSyntaxRow{}
        \roleDimension
        \figSyntaxSpaceItem
        \roleTime
        \figSyntaxSpaceNextCategoryMoreSpace
        \figSyntaxRowLabel{Param. Types}{\varType}
        \tDataTarget
        \figSyntaxLineBreak
        \figSyntaxRow{}
        \tMultiDataTarget
        \figSyntaxLineBreak
        \figSyntaxRow{}
        \tTagAndConfig{String}
        %
        \figSyntaxSpaceItem
        \tTagAndConfig{Number}
        \figSyntaxLineBreak
        \figSyntaxRow{}
        \tTagAndConfig{Boolean}
        \figSyntaxSpaceItem
        %
        \tTagAndConfig{Enum}
        \figSyntaxLineBreak
        \figSyntaxRow{}
        \tTagAndConfig{Text}
        %
        \figSyntaxSpaceItem
        \tTagAndConfig{Section}
        \figSyntaxSpaceNextCategoryMoreSpace
        \figSyntaxRowLabel{Spec. Lang.}{\varSpecLang}
        \specLangVega
        \figSyntaxSpaceItem
        \specLangVegaLite
        \figSyntaxSpaceItem
        \specLangAtom
        \figSyntaxLineBreak
        \figSyntaxRow{}
        %
        \specLangTable
        \figSyntaxSpaceItem
        \cdots
        %
        \figSyntaxSpaceNextCategoryMoreSpace
        \figSyntaxRowLabel{Views}{\varView}
        \textrm{(}\varSpecLang\textrm{-language specific rendering)}
        \figSyntaxEnd
    \end{minipage}%
    \begin{minipage}{0.5\textwidth}
        \ifonecol 
            \small
        \fi
        \vspace{-1em}
        %
        \judgementHead
        {Template Application}
        {
            \applyTemplate
            {\varTemplate}
            {\varSettings}
            =
            \varView
        }

        \vsepRule

        $
            \hfill
            \templateComponentProj{\varTemplate}{lang} = \varSpecLang
            \hfill
            \templateComponentProj{\varTemplate}{func} =
            \templateFunction{\varDecls{\varVar_1}{\varType_1}}{\varExp}
            \hfill
            \varSettings = \settings{\varVar_1}{\varAtom_1}
            \hfill
        $
        %
        %
        %
        %
        \vsepRule
        \begin{center}
            \begin{tabular}{ccc}
                $
                    \overbrace{
                        \applySubst{\varSettings}{\varExp} = \varExp'
                    }^{\textrm{A. Substitute Arguments}}
                $ &
                $
                    \overbrace{
                        \evalIvy{\varExp'} = \varVal
                    }^{\textrm{B. Evaluate Conditionals}}
                $ &
                $
                    \overbrace{
                        \evalLang{\varSpecLang}{\varVal} = \varView
                        \vphantom{\varSettings'} 
                    }^{\textrm{C. Render Visualization}}
                $
            \end{tabular}
        \end{center}

        \vsepRule

        \judgementHead
        {Evaluation of JSON Expressions}
        {
            \vphantom{\varSettings'} 
            \evalIvy{\varExp} =
            \varVal
        }
        \begin{align}
            \text{\textit{Atomics\ }}\hspace{-0.15in}                       &   &
            \evalIvy{\varAtom }                                             & =
            \varAtom                                                              \\
            \text{\textit{Objects\ }}\hspace{-0.15in}                       &   &
            \evalIvy{\jRecordN{\varStr_1}{\varExp_1}{\varStr_n}{\varExp_n}} & =
            \cup_i\
            {\evalIvy{\jRecordBindSingle{\varStr_i}{\evalIvy{\varExp_i}}}}        \\
            \text{\textit{Lists\ }}\hspace{-0.15in}                         &   &
            \evalIvy{\jListN{\varExp_1}{\varExp_n}}                         & =
            \jListN{\evalIvy{\varExp_1}}{\evalIvy{\varExp_n}}
        \end{align}

        \vsepRule
        \judgementHead
        {Evaluation of Conditionals}
        {
            \vphantom{\varSettings'} 
            \evalIvy{\varExp} =
            \varVal
            \textrm{\ or\ } \bot
        }
        \vsepRule
        \begin{align}
            \evalIvy{\iteOptional{\varCond}{\varExp_1}{\varExp_2}} & =
            \makebox[0.50in][l]{\evalIvy{\varExp_1}}
            \textrm{ if } \evalJS{\varCond} = \texttt{true}            \\
            \evalIvy{\ite{\varCond}{\varExp_1}{\varExp_2}}         & =
            \makebox[0.50in][l]{\evalIvy{\varExp_2}}
            \textrm{ if } \evalJS{\varCond} = \texttt{false}           \\
            \evalIvy{\iteNoElse{\varCond}{\varExp_1}}              & =
            \makebox[0.50in][l]{$\bot$}
            \textrm{ if } \evalJS{\varCond} = \texttt{false}          
        \end{align}
        \vsepRule
        \judgementHead
        {Evaluation of Conditional Fields}
        {
            \vphantom{\varSettings'} 
            \evalIvy{\jRecordBindSingle{\varStr}{\varExp}} =
            \jRecordOne{\varStr}{\varExp'}
            \textrm{\ or\ } \emptyset
        }
        %
        %
        \begin{align}
            \makebox[1.04in][r]
            {\evalIvy{\jRecordBindSingle{\varStr}{\varExp}}} & =
            \makebox[0.70in][l]{$\emptyset$}
            \textrm{ if } \evalIvy{\varExp} = \bot               \\
            \evalIvy{\jRecordBindSingle{\varStr}{\varExp}}   & =
            \makebox[0.70in][l]{$\jRecordOne{\varStr}{\evalIvy{\varExp}}$}
            \textrm{ otherwise}
        \end{align}
    \end{minipage}
    \caption{
        The Ivy template language is composed of an abstract syntax grammar (left) and evaluation rules (right).
    }
    \Description[The formal syntax of our langugage]{The formal syntax of our langugage. The left side shows a series of types, while the right side shows how those types are applied.}
    \label{fig:lang}
    \label{fig:interp}
\end{figure*}

We now describe the design and implementation of \ivy{}.
First, in \secref{sec:narrative-walkthrough}, we introduce to the components of our system with a narrative walkthrough.
Next, in \secref{sec:lang-design}, we present a rigorous formulation of \emph{parameterized declarative templates}---a grammar-agnostic mechanism for abstracting JSON-based specifications---in a small programming language.
Then, in \secref{sec:ui}, we describe our UI design for selecting and instantiating templates, designed to obtain the benefits and familiar aesthetics of chart choosing and shelf building.
Finally, in \secref{sec:emergent}, we describe how the systematic application of templates enables beneficial forms of view exploration.

\subsection{Narrative Walkthrough}\label{sec:narrative-walkthrough}

We now introduce the technical components of our system with a narrative description of our hypothetical user Tabitha constructing a reusable template from an example.

She begins by loading a population dataset in \system{}. She then copies in the
``Aggregate Bar Chart''~\cite{aggregateBarChart} from the Vega-Lite documentation (\autoref{fig:bar-chart}b).
\autoref{fig:bar-chart}a shows the output visualization.

After pasting in the code into the \emph{template body} (via the Body tab of \figref{fig:annotated-interface}c), several automated suggestions are provided on how the data fields could be abstracted as \emph{template parameters}.
Clicking through the suggestions replaces the \jString{age} and \jString{people} data fields (\autoref{fig:bar-chart}b, lines 8,10) with two new \verb+DataTarget+ parameters (\autoref{fig:bar-chart}c, lines 7,13).
Tabitha renames the generated parameters to \jParam{xDim} and \jParam{yDim} which automatically replaces their uses (enclosed by ``escape'' brackets) with \jParam{[xDim]} and \jParam{[yDim]} (\autoref{fig:bar-chart}c, lines 23-24). She uses the settings popover (\figref{fig:edit-mode}) to specify their allowed data roles---she could have also done so using the Params text box (\figref{fig:annotated-interface}c).
These configurations result in a shelf builder style user interface which she uses to explore her data set. She then uses these shelves to specify values for the \verb+xDim+ and \verb+yDim+ parameters (\autoref{fig:bar-chart}d). These \emph{value settings} are then \emph{applied} to the \emph{template body} to produce a JSON specification, which is then transformed into a visualization by a language-specific rendering function, in this case, by Vega-Lite.

Next, she wants to chart populations for different years, so
she replaces the constant value \verb+2000+ (\autoref{fig:bar-chart}b, line 5) with \verb+[year]+ (\autoref{fig:bar-chart}c, line 20),
referring to a new template parameter that abstracts over the choice of year
(lines 9-12).
Intending to apply this template only to datasets with decennial measurements, Tabitha specifies
that \verb+year+ should be chosen from among a new \verb+Enum+ type, called \jString{allowedValues},
comprising census years (\autoref{fig:bar-chart}c, line 11).
(She chooses not to abstract the filtering predicate, which would have resulted in a both more general and more complex template.)

Then, Tabitha considers whether and how to sort the bars.
Knowing the appropriate Vega-Lite option, she adds a \jField{sort} field with a conditional to the template body (\autoref{fig:bar-chart}c, line 23).
She then creates a new template parameter, \verb+sort+ (line 8), and configures it such that if \verb+sort+ is set to \verb+true+, then the \jField{sort} value in the resulting Vega-Lite specification is set to \jString{-x}, which, in rendered Vega-Lite, sorts the bars in order of increasing value.
If \verb+sort+ is set to \verb+false+, then the resulting spec contains
no \jField{sort} field---as in the original specification
(\autoref{fig:bar-chart}b).

Lastly, Tabitha wants to enable control of the color of the bars in her chart. To do so, she first defines a \verb+color+ parameter (\autoref{fig:bar-chart}c, line 14), and then adds a reference to this parameter on line (\autoref{fig:bar-chart}c, line 21). This allows her, or other users of this template, to pick between colors preferred by her organization.

Her template now satisfactorily prepared, she is ready to explore her dataset by specifying parameter values through the GUI (as in \autoref{fig:bar-chart}d). In addition to specifying individual value settings, she can explore multiple options simultaneously using \emph{fan out} (\autoref{fig:bar-chart}e), which juxtaposes multiple views to consider (\secref{sec:ui-fan-out}).

\subsection{Template Language Design}\label{sec:lang-design}

\newcommand{\templateComponentUnderline}[1]
{\templateComponent{#1}}

Templates provide a simple set of abstractions over JSON-based grammars. Put simply, a template is a function specified in a superset of JSON, which includes variables and simple control flow operators, that when applied to arguments produces a chart in a particular visualization grammar. Templates are grammar-agnostic as they abstract arbitrary JSON specifications.
We provide a full description of templates to highlight exactly how they make declarative grammar specifications reusable (\goal{4})---by combining multiple specifications into a single template---and easier to use (\goal{1})---by demarcating the arguments for manipulation.

Formally, we define a \emph{template}~$\varTemplate$ as a function that applies a set of $N$ \textbf{\emph{template parameters}}, $\varVar_i$ with type $\varType_i$, to a \textbf{\emph{template body}}~$\varExp$ to generate a \emph{view} $\varView$.
In addition to the \templateComponentUnderline{func}tion itself, a template is a record that defines:
the output JSON specification \templateComponentUnderline{lang}uage $\varSpecLang$ of the template body,
\templateComponentUnderline{metadata},
and a list of zero or more template-specific constant \templateComponentUnderline{symbols} that the body may refer to (\secref{sec:template-parameters}).
Our simplified model of JSON~\cite{ecma_json_2017} values $\varVal$
comprises \emph{atomic values}~$\varAtom$---literal strings
$\varStr$ of type \tTag{String},
numbers $\varNum$ of \tTag{Number},
and booleans $\varBool$ of type \tTag{Boolean}---records of string-value
pairs, and lists of values.
The abstract syntax of templates is defined in \figref{fig:lang}.

\subsubsec{Template Bodies}
\label{sec:lang-body}
Beyond JSON literals, a template body, or \emph{expression} $\varExp$, may employ two basic programming constructs.
A \emph{variable} refers to a template function parameter $\varVar$ or a template symbol $\varSymbol$, wherever a JSON literal value might normally appear.
A \emph{conditional expression}, written $\iteOptional{\varCond}{\varExp_1}{\varExp_2}$, is dependent on the result of \emph{predicate expression} $\varCond$, which is a ``raw'' JavaScript code string that evaluates to a boolean value.
The else-branch is optional, which supports optionally-defined fields, per \secref{sec:lang-interpret}.

\figref{fig:bar-chart}c shows the concrete syntax for variables and conditionals
in \ivy{}, which implements the abstract syntax of \figref{fig:lang}.
Variables are ``escaped'' with square brackets (\eg{}~``\verb+[year]+'').
Following similar support in other tools~\cite{mongodb_inc_mongodb_nodate}, conditionals are written as shown in \figref{fig:bar-chart}c line 23.
Full-featured languages typically provide more extensive abstraction mechanisms, but---as we show in \secref{sec:eval}---even just variables and conditionals suffice for a wide variety of use cases.

\subsubsec{Template Parameters}\label{sec:template-parameters}
The $N$ parameter declarations $\varVar_i$ with types $\varType_i$ serve two purposes:
to abstract over data fields and stylistic choices in the definition of a
visualization and
to define GUI elements that allow users to specify argument values for these parameters.
For example, a ``switch'' widget is drawn for each parameter of type \tTag{Boolean},
a dropdown menu for each \tTag{Enum}, and
a Tableau-style ``shelf'' for each \tTag{DataTarget}
(\figref{fig:bar-chart}d).
\figref{fig:lang} defines several parameter types $\varType$.
The data target types \tTag{DataTarget} and \tTag{MultiDataTarget} range
over data columns (identified with strings) in the current dataset as
well as template-specific symbols (also encoded with strings in our
implementation).
Symbols are template-specific values that can be used to instantiate parameters.
For example, \ivyPolestar{} defines a \jParam{count} symbol that induces a targeted channel to take on a count aggregation.

Each type $\varType_i$ contains type-specific arguments $\varConfig_i$,
for example, the minimum and maximum allowed values for a \tTag{Number} or the
values comprising an \tTag{Enum}.
The data parameter types---\tTag{DataTarget} and \tTag{MultiDataTarget}---carry a configuration field to define a \emph{data role} $\varRole$, discussed further in \secref{sec:ui}.
Each type $\varType_i$ also contains an optional boolean expression $\varCond_i$
that determines whether the GUI should display controls for that parameter $\varVar_i$---%
for example, displaying a sort direction widget only when a boolean \jParam{sort} parameter is set to \verb+true+.
\figref{fig:lang} elides details of these type-specific arguments.

As parameters are arguments to a function, they must also have values.
We refer to a user's choice of argument values $\varAtom_i$ as \emph{settings} $\varSettings$.
Settings are typically set using the GUI, but can also be specified as text (\figref{fig:annotated-interface}c, ``Settings'' tab).

\subsubsec{Evaluating Templates to Visualizations}
\label{sec:lang-interpret}
Finally, we produce visualizations
by applying the function described by template $\varTemplate$ to the parameter settings $\varSettings$,
following the three steps found in \figref{fig:interp} A-C.
\textbf{First}~(\figref{fig:interp}A), the argument values $\varAtom_i$ are substituted for the parameters $\varVar_i$ in the template body~$\varExp$ using straightforward substitution.
For example, this transforms a snippet of the body in \figref{fig:bar-chart}c, \verb+{"y": {"field": "[yDim]"}, ...}+
with settings \verb+{"yDim": "age", ...}+ to be \verb+{"y": {"field": "age", ...}}+.

\textbf{Second}~(\figref{fig:interp}B), the resulting expression is evaluated to produce a JSON value, transforming any conditionals into JSON values.
This is done by a straightforward recursive traversal, following Eqns (1)-(3).
Whenever a conditional is encountered, it is executed by evaluating the JS-based predicate ($\evalJS{\cdot}$ referring to JS evaluation), and then replacing the conditional with either the corresponding then- or else-branch as appropriate.
Eqns (4) and (5) handle the two usual cases for conditionals.
Eqn (6) handles the case when the predicate evaluates to \verb+false+ but no else-branch is provided; returning a $\bot$ value, indicating that the conditional is to be deleted.
Eqns (7) and (8) explicitly provide a mechanism to delete conditionals that return $\bot$ in record values.
This is how, for example, \figref{fig:bar-chart}c line 23 optionally defines a \jField{sort} field.
Our implementation generalizes this rule to arrays by viewing them as numerically indexed records.

The \textbf{third}~(\figref{fig:interp}C) and final step is to use the generated JSON to render a visualization. This is done by, for a specific visualization language $\varSpecLang$, using the corresponding rendering function
$\evalLang{\varSpecLang}{\varVal}$ to interpret the JSON value and render the
resulting view---typically as an HTML Canvas or SVG.
Our current implementation supports four languages $\varSpecLang$, namely, Vega,
Vega-Lite, Park \etals{} Atom~\cite{park_atom_2018}, and a simple table language.
However, as we discuss in \secref{sec:language-extensibility}, \ivy{}'s language support is designed to be extensible.

\subsection{User Interface Design}
\label{sec:ui}

Equipped with the notion of templates, we next describe the user interface design of \ivy{}.
As shown in \figref{fig:annotated-interface}, the application consists of two panes, one for chart editing and another for chart viewing.
The chart editing pane contains a data column filled with Tableau-style ``pills'' representing data columns, and an encoding column with ``shelves'' for those pills to be placed upon.
This encoding column can be used to instantiate (\ie{}~provide arguments for) the parameters of the template, or to edit the GUI of the current template.
The editing pane also includes a code editor which can manipulate the current template and UI state textually.
Per Saket \etal{}~\cite{saket_liger_2019}, we facilitate multimodal interaction by tightly synchronizing these views, such that changes in one modality are instantly reflected in the other.

Below, we describe how \ivy{} supports the \emph{creation}, \emph{selection}, and \emph{application} of templates to produce charts.

\subsubsec{Template Selection as Chart Choosing}
The root of \ivy{} is a template gallery, which aims to achieve Satyanarayan \etals~ \cite{satyanarayan_declarative_2014} vision for a system ``allowing users to browse through designs for inspiration, or adapt them for their own visualizations.''
This avoids the blank canvas problem \cite{satyanarayan_critical_2020} and supports ease of use~(\goal{1}).

The gallery is populated with a library of system-provided templates, as well as templates created by \ivy{} users (stored on a communal server).
Simpler templates allow users to jump quickly to familiar visual forms (such as line charts or bar charts), while more sophisticated templates privilege thinking with their data~\cite{romat2019activeink} (as in \ivyPolestar{}).
The gallery is present both as a homepage for the application, independent from the visualization editor, as well as an intermediate view while creating new view tabs.

Each template is accompanied by a set of user-defined examples, namely, settings chosen by users to instantiate the template, with data bindings and output renderings with respect to a collection of predefined datasets.
These examples serve as ``crowd-sourced documentation'' for how individual templates operate.
Furthermore, this adds an element of opportunistic programming: to create templates, users can borrow small snippets---such as a well-formatted list of color schemes---and use them in their own creations.

\subsubsection{Template Application via Shelf Building}

After selecting a template and uploading a dataset of interest, the user is presented with a shelf builder-style GUI for setting the template parameters and specifying basic data filters.
We choose this GUI design seeking to exploit the same affordances that drive the explorability~(\goal{2}) of shelf builders.
Specifically, our design closely follows that of the Polestar shelf builder system, which Wongsuphasawat \etal{}~\cite{wongsuphasawat_voyager_2015, wongsuphasawat_voyager_2017} constructed as a simulacra of Tableau to serve as a baseline comparison in the development of their recommendation-based exploration systems.
While emulating Polestar is a relatively small threshold to overcome in the context of visualization systems in general, it demonstrates the promise of our template-based approach.
Systems such as Tableau or PowerBI possess features that---although substantially larger and more complex---are not substantially \emph{different} than those in Polestar.
To select data parameters (\ie{}~\tTag{DataTarget} or \tTag{MultiDataTarget}) of interest, users drag-and-drop from a list of data fields, color-coded according to their \emph{data roles}, onto encoding shelves, as in \figref{fig:annotated-interface}a, d.
Following prior work~\cite{stolte_polaris_2002,agarwal2019viswall,wongsuphasawat_voyager_2015}
roles include $\roleMeasure${ } (quantitative fields), $\roleDimension${ }
(nominal or ordinal fields), and $\roleTime${ } (temporal fields).
When a dataset is loaded, we make heuristic guesses about the role for each column, which the user can later modify.
We use roles in \ivy{} to construct a naive automatic \textit{Add to Shelf} feature (akin to Tableau's Add to Sheet \cite{mackinlay_show_2007}), except ours is simply based on order and data role.
If a template has three \tTag{DataTarget}s, the first of which allows only a
$\roleMeasure${ } while the latter two allow anything, clicking \textit{Add to
    Shelf} on a $\roleDimension${ } will add it to the second parameter.

\begin{figure}[t]
    \begin{minipage}{\oneColFigWidth}
        \centering
        \includegraphics[width=\linewidth]{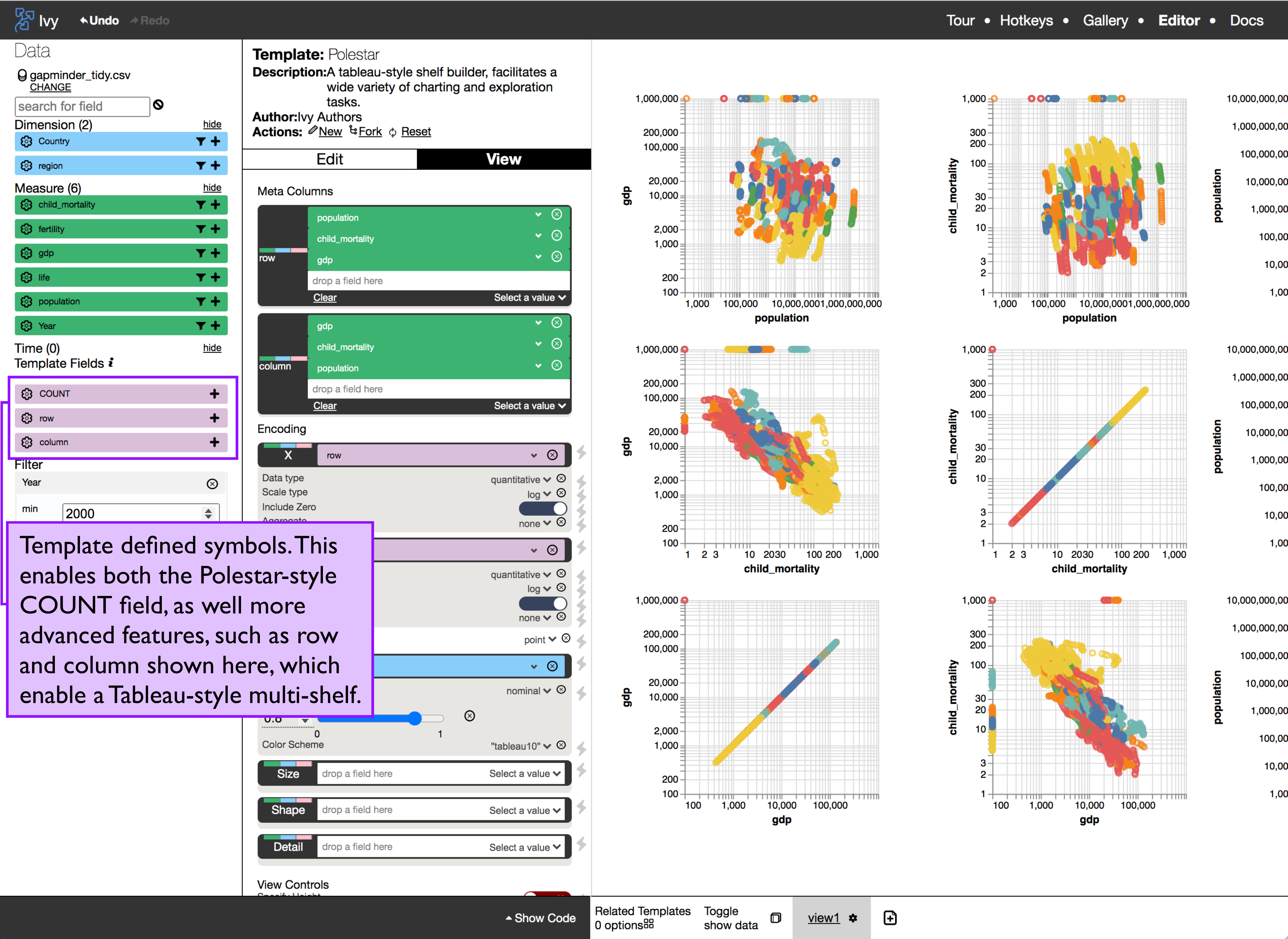}
        \caption{Our \ivyPolestar{} template reproduces and extends the functionality of the Polestar shelf builder. Here template specific
            symbols are used to make a SPLOM.}
        \label{fig:polestar}
        \Description[A view of the ivyPolestar template]{The Ivy web application set to the Ivy Polestar template. A log-log scatter plot matrix is currently formed by placing row and column pill onto the x and y axes, and by placing a series of data columns into multi shelves at the top of the template. The splom shows the gapminder dataset for gdp, child mortality, and population.}
    \end{minipage}
\end{figure}

In addition to the visual aesthetics of Polestar (and hence that of Tableau), we also emulate the \emph{functionality} of its shelf-building interface through a ``default'' library template called \ivyPolestar{} (shown in \figref{fig:polestar}).
The only features not replicated are the Automatic Mark Type---implementation of which, though possible in \ivy{}, was beyond the scope of the paper---and the chart bookmarks---which we replaced with a notion of view tabs.

A notable feature in Polestar is a \emph{count} symbol that is visually similar to normal data fields and induces count aggregations on channels without a selected data column.
To model this feature, we define a template-specific symbol (\cf{}~\secref{sec:template-parameters}), \jParam{count}, which \ivyPolestar{} uses to implement the corresponding functionality.
\ivyPolestar{} introduces two additional symbols, \jParam{row} and \jParam{column}, that enable faceting by data columns in the manner of Tableau's multi-shelves.
Placing \jParam{row} or \jParam{column} symbols on any shelf creates a \tTag{MultiDataTarget} which acts as a wrapper around Vega-Lite's juxtaposition operators \cite{satyanarayan_vega-lite_2016}.
\figref{fig:polestar} highlights this feature.

\subsubsection{Template Creation and Text Editing}\label{sec:text-editing}

\begin{figure}[t]
    \begin{minipage}{\oneColFigWidth}
        \centering
        \includegraphics[width=\linewidth]{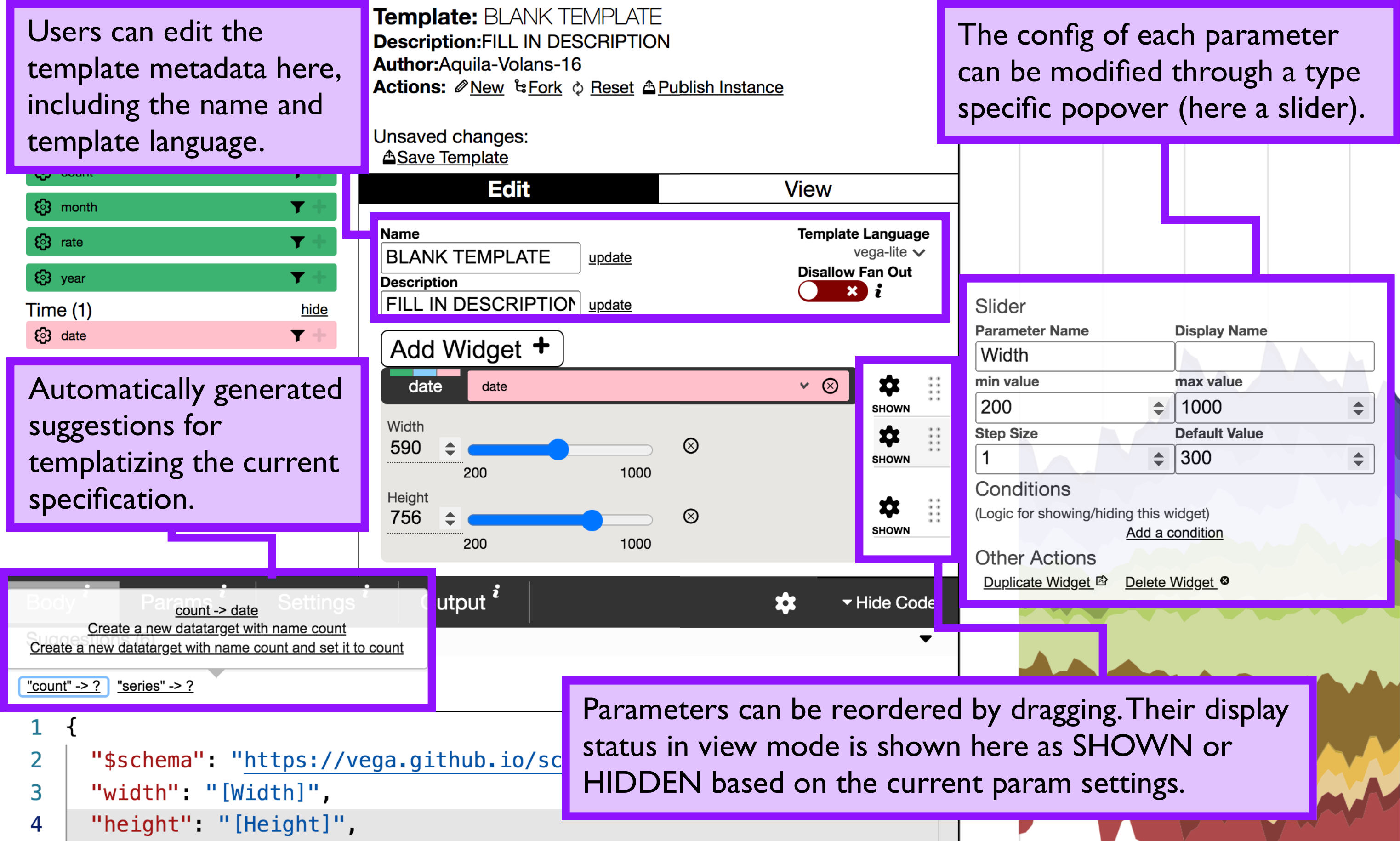}
        \vspace{-0.15in}
        \caption{In edit mode, users can modify templates both through the shelf builder GUI and the textual code pane.
        }
        \label{fig:edit-mode}
        \Description[An annotated view of the edit mode GUI]{An annotated view of the edit mode GUI in the Ivy web application. The current chart is a streamgraph of unemployment-across-industries, colored by industry type. In the code area a suggestion appears that offers several suggestions on how to templatize the code. It appears that a user is in the process of creating a new template.}
    \end{minipage}
\end{figure}

Templates can be created or modified in two ways, either by modifying the textual representation (\figref{fig:bar-chart}c) or through GUI interactions (\figref{fig:edit-mode}).
The textual representation facilitates both small tweaks (\secref{sec:novice}), as well as creating new templates.
For instance, users may copy code snippets found online---such as in language documentation or Stack Overflow---and templatize them to suit their task (\secref{sec:expert}).
Templates can also be created by ``freezing'' and refining the GUI state when interacting with an existing template.
For example, a user might apply a full-featured template, such as \ivyPolestar{}, to construct something resembling their desired chart, fork the text output as a new template (as in \figref{fig:annotated-interface}d), and then provide fine textual grained updates.
As discussed previously, exposing textual representation to the end user furthers our flexibility (\goal{3}) and reusability (\goal{4}) goals.

To ease the construction of templates, \ivy{} uses domain-specific pattern matching and rewrite rules to suggest potential transformations to users.
For instance, if a user were to find a chart in the Vega documentation that they wanted to copy, they would simply start a new template and paste the code into \ivy{}.
The code pane then suggests ways to transform the code.
For example, if a value in a Vega-Lite spec (such as in \figref{fig:bar-chart}b) is used where a data reference is expected (e.g. \texttt{"field": "age"}), then \ivy{} suggests swapping \texttt{"age"} with a reference to a new parameter.
\figref{fig:edit-mode} shows an example of such rules.
Rules are defined by Ivy developers, rather than \ivy{} users.
We implemented rewrite rules for Vega-Lite, Vega, and Atom.

\subsection{Template-Based View Search}\label{sec:emergent}

The systematic formulation and application of templates allows us to emulate recommendation and exploration (\goal{2}) features found in a variety of existing charting systems as a consequence of our design.
Here we highlight two such features that follow naturally from the use of templates:
one arises by fixing the arguments and varying the template, and
the other by fixing the choice of template and varying the arguments.

\subsubsection{Extensible Recommendation with Catalog Search}
\label{section:catalog}
\label{sec:ui-catalog-search}

The heterogeneity of user needs is often addressed in chart choosers by offering large and often diverse sets of charting options \cite{mackinlay_show_2007}, which can be intimidating or difficult to utilize due to their volume.

The gallery in \ivy{} is equipped with \emph{catalog search}, which allows users to search across the set of available templates based on compatibility with a set of specified columns of interest---specifically, by using a simple type-compatibility algorithm that compares template parameter types with the data roles of selected columns.
This feature exists both in the gallery---where it acts as a type-based search mechanism---as well as ambiently throughout the system in the related templates tab (\figref{fig:annotated-interface}f)---where it acts as a simple alternative recommendation system.
The current template is compared with each other template, yielding a \emph{partial match}, a \emph{complete match}, or \emph{no match}.
A partial match occurs when the selected data columns can be mapped to a template's parameters. A match is complete if all required parameters are mapped.
A complete match can translate the current selection without additional specification.
This heuristic is further detailed in the appendix.
When a match is selected, the current columns are mapped onto the new template, using a similar mechanism as our \emph{Add to Shelf}, and the resulting chart is shown immediately.

\begin{figure}[t]
    \begin{minipage}{\oneColFigWidth}
        \centering
        \includegraphics[width=\linewidth]{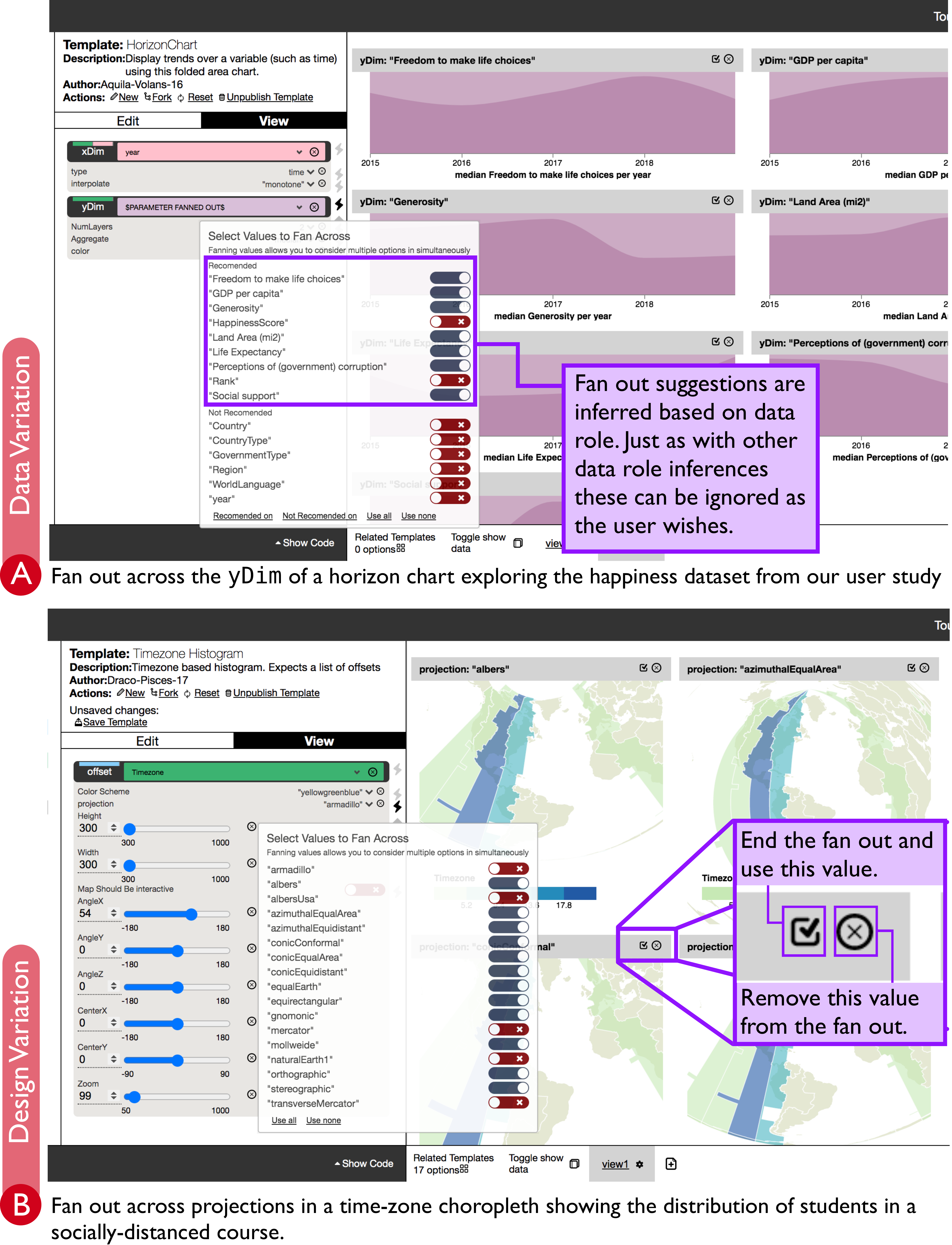}
        \vspace{-0.2in}
        \caption{
            Users rapidly explore design and data alternatives via fan outs.
            After selecting parameter values of interest, they are shown all variations simultaneously.
        }
        \label{fig:fan-out}
        \Description[Two views of the fan out technique]{Two views of the fan out technique juxtaposed vertically. A) is labeled "Data Variation" and shows a set of purple horizon charts that have the same x axis and differing y-axis. The fanout menu for the y axis is open and is annotated. The data is the world happiness dataset. B) is labeled "Design Variation" and shows a set of choropleths varied by map projection. The fan out menu for the projection is open and various projections are switched on and off. There is a call out box over the center of the choropleths highlighting controls present in the fan out.}
    \end{minipage}
\end{figure}

\subsubsection{Exploring Encoding Variation through Fan Outs}
\label{sec:ui-fan-out}

Comparisons in visual analytics are often made temporally, requiring the analyst to hold mental reference to each of the values under consideration.
To reduce this cognitive burden, \ivy{} users can \textit{fan out} a template by applying multiple settings and rendering their output simultaneously.
\figref{fig:bar-chart}e and \figref{fig:fan-out} display examples of this interaction.

To begin a fan out a user specifies sets of values that they wish to compare for each parameter of concern.
This can be applied to both design and data parameters.
For data parameters we provide suggestions of appropriate values based on the inferred data role and specified parameter role (\figref{fig:fan-out}a).
We then compute a cartesian product of these sets and render a separate instance of the current template for each combination of values.
Users then browse the resulting gallery, and can modify all of the combinations at once through the shelf-building UI, as well as remove values or select a combination to view in isolation (thus collapsing the fan out).

This approach allows users a low-stakes way to consider alternative chart configurations and rapidly explore the space of available design and data parameters.

\section{Evaluation}
\label{sec:eval}

In the previous section, we discussed how our system design integrates and augments UI capabilities provided by existing chart choosers, shelf builders, and text-based editors.
Here, we assess how well our template-based approach may work in practice by considering two questions:
\textbf{\emph{Do templates facilitate organization and reuse of existing visualizations?}}
(\secref{sec:vega-lite-templates}), and
\textbf{\emph{Is Ivy's multimodal UI approachable by real users?}}
(\secref{sec:usability-study})

\subsection{Templates for Existing Visualizations}\label{sec:vega-lite-templates}

We considered two chart corpora, the example gallery of Vega-Lite~\cite{vegaLiteDocs} and the chart chooser found in Google Sheets~\cite{googleSheetsChartChooser}, looking for opportunities to factor related visualizations into templates, which yielded 3.5x and 1.8x \emph{compression ratios},
respectively.
Compression is the number of examples constructible by a given template.
The reported average compression are computed by
\begin{equation}
    \frac{|\text{Examples}| - |\text{Excluded by Data Model}|} {|\text{Templates}|}
\end{equation}
In an example gallery, a larger ratio means less duplicated code (text) among examples. In a chart chooser, it means fewer chart forms with possibly more parameters.
These results demonstrate that the simple template abstraction mechanisms enhance the flexibility~(\goal{3}) of existing declarative grammars while improving their reusability (\goal{4}) by serving a variety of use cases.

While even just the basic abstraction mechanisms in templates---namely, variables and conditionals---are sufficient for merging all examples of each corpus into single templates,
we strove to construct templates that are factored in reasonable ways, for example, by not depending on datasets having particular columns or features, and by considering what might plausibly be found in a chart chooser.
This process prompted a number of minor system improvements, such as simplifying the conditional syntax and improving error handling, and suggested directions for future work on Ivy, such as a more sophisticated data model.
Additional details of this analysis can be found in the appendix.

\subsubsection{Vega-Lite Example Gallery}
\label{sec:vega-lite}

\begin{figure}[t]
    \begin{minipage}{\oneColFigWidth}
        \centering
        \includegraphics[width=\doubleColFigWidth]{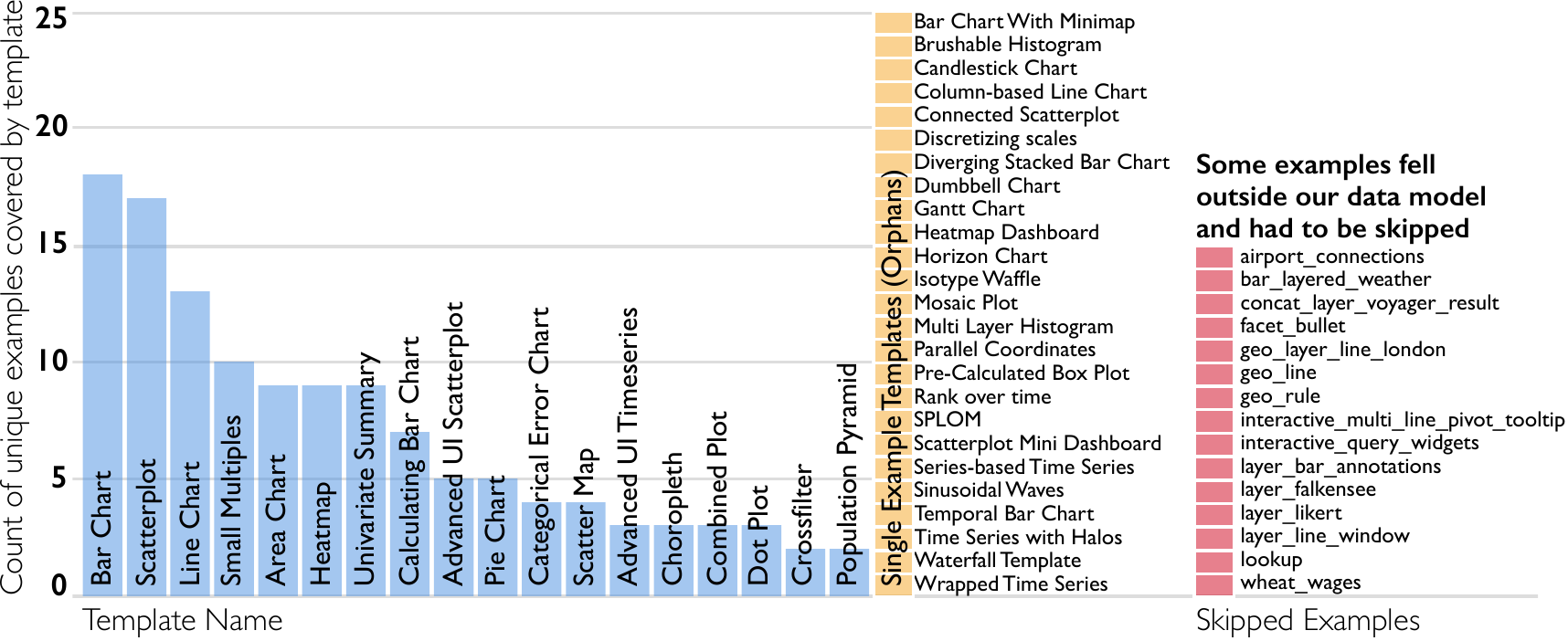}
        \caption{
            We reproduced the 166 examples found in the Vega-Lite gallery using 43 templates.
            Some examples had to be excluded (right) as they fell outside our data model, usually because they utilize multiple data sources, which \ivy{} does not currently support.
        }
        \label{fig:templatization-report}
        \Description[A bar chart of the templatization of the Vega-Lite gallery]{A bar chart of the templatization of the Vega-Lite gallery. The left side shows a set of blue columns with template names superimposed over them indicating the number of examples covered by that template. The middle column shows a series of yellow boxes superimposed with "Single Example Templates (Orphans)" and their name on their right. Off to the side is a column of red boxes show examples that had to be skipped.}
    \end{minipage}
\end{figure}

This gallery consists of 166 distinct specifications,
reflecting years of iterative refinement and development by the Vega-Lite community.
These examples highlight a breadth and depth of features exposed by the library, thus forming an ideal testbed for the utility of templates.
We recreated this corpus with 43 templates, skipping 14 examples due to incompatibility with our data model (a single, flat input data table) for a 3.5x compression.
\figref{fig:templatization-report} reports the frequency of each template.

We aimed to capture both the core content of each example (the major feature being demonstrated) as well as the resulting image. However, this was not always possible.
We allowed minor text modifications and the inclusion of properties not present in the original example, as long as they did not affect the core of the example (such as styling).
It was sometimes necessary to forgo or modify some examples to accommodate our current implementation. For instance, a scatter-map of zip-code centroids colored by their first digit~\cite{centroidExample} needed to be modified because Ivy currently lacks a custom calculation feature.
As a way to guide this design, we strove to maximize the \emph{concatenation ratio}, or the ratio between the size of a template body and that of the concatenated examples it captures, which is computed for a particular measure of size $\gamma$,
\begin{equation}\label{eqn:concatRatio}
    \frac{\sum_{x \in \text{covered examples}} \gamma(x)}{\gamma(\text{template})}
\end{equation}
The resulting templates have an average lines of code concatenation ratio of 1.48x (a proxy for simplification) and an average abstract syntax tree concatenation ratio of 1.80x (a proxy for UI complexity minimization).
In conjunction with the average example compression ratio of 3.5x, this suggests that our templatizations
are better than merely concatenating the examples.

\subsubsection{Google Sheets Chart Chooser}\label{sec:google-sheets}

The set of charts found in this chooser consists of 32 distinct chart forms.
We focused on Google Sheets as a representative chart chooser because it possesses a similar set of chart options as other choosers, such as Excel and LibreOffice.
We reproduced 29 of these charts through 16 templates for a 1.8x compression factor. We skipped 3 examples for the following reasons:
``3D Pie'' charts require a 3D visualization grammar,
``Org. Charts'' fall outside of our tabular data model, and
``Timelines'' require annotations, and therein modifications to the underlying dataset, which is not currently supported in \ivy{}.

\subsection{Approachability Study}\label{sec:usability-study}

We conducted a study in order to evaluate whether the multimodal user interface in \ivy{} is approachable by real users.
During pre-study pilot sessions, we found that users who were more familiar with Vega-Lite learned more quickly how to integrate Ivy's full capabilities to address tasks than those unfamiliar with Vega-Lite.
Therefore, we aimed to recruit users familiar with these paradigms in order to evaluate the approachability of interacting with multiple charting modalities, rather than that of the underlying grammars.
The target expertise for our study can be characterized roughly as between that of the hypothetical users from \secref{sec:axel-tabitha}.

\subsubsection{Participants}

We solicited 5 participants
from a recent visualization course in a computational public policy masters program, all of whom are now employed as professional data analysts.
Based on their participation in the course, we were confident
that these now-graduated students were familiar with Vega-Lite and Tableau.
Participants, denoted P1 through P5, self-reported a mean familiarity with Vega/Vega-Lite of $\mu=2.4$ on 5-point Likert scale, and $\mu=3.2$ with visual analytics tools  (5 indicating high familiarity).
Participants were students at the same institution as the experimenters---specifically, in the context of a student-instructor relationship---thus constituting a source of potential bias.

\subsubsection{Methods}

Following a brief introduction to \ivy{} through in-application documentation, participants completed a series of 8 tasks that covered data exploration, chart specification, and template construction problems (additional details in the appendix).
For instance, one task asked participants to templatize a particular Vega-Lite-based box plot~\cite{boxPlotExample}, while another asked them to find the number of countries in a multi-year version of the World Happiness Report~\cite{helliwell2012world}.
We focused on these tasks because they are similar to tasks supported in related systems and (with the exception of template building) were addressable with any of the modalities individually or in conjunction.
Correctness was judged by comparison with a solution set prepared prior to the study.
We engaged subjects in an informal think-aloud protocol during the session, which led to a number of the reported observations in the next section.
Subjects were then asked to fill out an exit survey on the usability of Ivy and templates.
Sessions were held over video conference software and lasted a mean of $\mu=95$ minutes, although 2 hours were allotted. Participants received a \$50 Amazon gift certificate for their work.

\subsubsection{Results}

All participants were able to complete all tasks within the allotted time, although all users required some assistance at various stages, typically due to implementation bugs or learnability hurdles.
A variety of strategies were used to accomplish the data exploration tasks, some using only shelf building, some only chart choosing, or a mixture of both.
All users were able to complete the template construction tasks and then use the templates to address data exploration tasks.
We believe this indicates that real users, once acclimated, are able to produce non-trivial templates to accomplish varying goals.

Ths system was generally seen as usable. Participants mostly agreed with the statement ``I think that I would like to use this system frequently" ($\mu=4.4$ on a 5-point Likert scale), and gave a mean system usability score of $\mu=68.0$---describable as being between ``OK'' and ``Good''~\cite{bangor2009determining}.
More critical than users' perception of the usability, which may have been positively biased, is the demonstration they were able to navigate the system and use the interlocking modalities to achieve various tasks.
While our study did not cover some concepts in our system---such as conditionals and catalog search---it demonstrated that users can approach and utilize the various UI modalities in Ivy.
Participants sometimes had to be directed to use certain unfamiliar features, such as the templatization suggestions and fan outs.
However, once familiar, users tended to continue to use those features.

Participants were enthusiastic about mixing code and graphical specification. P5 commented that the combination \emph{``feels more useful than just coding''}.
P4 noted that their organization had recently switched a major Tableau-based dashboard to a Shiny-based dashboard because of a lack of precise data controls in Tableau. In contrast to the push for visual analytics tools to completely shed ties to text (embracing a ``no-code'' approach),
we believe this suggests that systems that straddle the boundary between code and GUI specification can offer a valuable mixture of affordances that support real use cases.

\subsubsection{Limitations}

We observed that \ivy{} held some challenges for participants.
Participants sometimes struggled to understand what the templatization suggestions would do, indicating a {closeness of mapping}~\cite{green1989cognitive} failure.
P1 and P2 suggested that visual design could be improved to aid in feature discovery.
While some participants (P2, P4) agreed that templates are good for creating rapidly reusable charts, P2 noted that templates are \emph{``not very portable.''}
This could be addressed by embedding \ivy{} in tools like Shiny~\cite{shiny} or Jupyter~\cite{jupyterWidgets}.

The learnability of the system and mixed-modality remained a primary difficulty. Users typically struggled to figure out how to bridge the gap between text and GUI at first, however, by the end of the session, all users were competent in both regimes.
For instance, most subjects iteratively refined their solutions to the box plot task, modifying both code and GUI values to address developing hypotheses.
P3 noted that the system required a non-trivial level of computational and visualization literacy.
Fortunately, users could seemingly bootstrap their knowledge to overcome these hurdles.
P4 noted that the integration between the \emph{``code body and the point-and-clickable GUI is really tight and also good for reinforcing learning''} and that \emph{``If you know how to do something in one form, you can do it and watch how it changes the other side of the tool.''}

The scope of this study was small. We merely sought to demonstrate that real users of similar systems could approach the mixed-modality UI found in Ivy.
While these results suggest that this combination is promising, further investigation is required to understand its utility in the context of more developed interfaces.
Once our system reaches maturity, we intend to conduct a study comparing it with standard analytics tools, such as Tableau or Excel.

\section{Discussion}
\label{sec:discussion}

\newcommand{\va}
{visual analytics}

In this paper, we described how \textit{parameterized declarative templates}---a typed abstraction layer over JSON specifications---can serve as a basis for a multimodal UI to create and explore visualizations.
\ivy{}-style templates may help in the organization and reuse (\goal{4}) of existing visualization corpora (per \secref{sec:vega-lite-templates}).
Vega and Vega-Lite have garnered ample popularity, and new declarative visualization grammars are being actively developed~\cite{li2021P6,kim2021Gemini,wongsuphasawat2020encodable}.
As the availability and use of these grammars continues to proliferate, there is opportunity for shared platforms and tooling between languages, which we explore in our grammar-agnostic templates.

The integration of features in our prototype appears to be accessible to users with modest experience in both \va{} systems and Vega/Vega-Lite (per \secref{sec:usability-study}).
Users were able to make effective use of affordances for exploration found in our shelf building UI and fan out (\goal{2}), and were able to utilize the capability of templates to improve the ease of use (\goal{1}) and reuse (\goal{4}) of declarative chart specifications while maintaining their flexibility (\goal{3}).

We believe that this multimodal approach has value for a variety of use cases.
Exposing a connection between GUI and programmatic API may enable analysts to self-serve their chart creation needs.
If a particular chart form is not available (but is constructible by one of the supported grammars) then they can create it for themselves, rather than requiring reliance on engineering resources.
This connection between text and GUI appears to help users learn and comprehend JSON-based charting grammars, which may be unfamiliar or difficult to understand.
The repeatable customization found in templates might also, for example, enable practitioners (e.g. data journalists) to explore designs in a structured manner that does not violate their organization's visual identity.

\subsection{Limitations and Future Work}

The version of multimodal visual analytics found in our prototype has its share of limitations.
The strength of each modality in \ivy{} is only as good as its implementation, which can render artificial barriers between what users expect and what is supported (\eg{} P3 expected a pivot table).
And while Ivy encompasses chart choosing, shelf building, and textual specification,
it does so at the cost of an increased learning curve.
However, we believe that this difficulty is not endemic to multimodal systems, and that through attentive design the experience of using the system can be made easier.

In making our template-based approach more viable for practical use, it is easy to imagine a variety of system improvements---such as additional template parameter types (e.g. color schemes or inline data fields), drag-and-drop interactions for refactoring and abstracting specifications~(\eg{}~\citep{DNDRefactoring,sns-deuce}), as well as enriching the ways in which changes made in one modality are reflected and explained in the others.
Beyond these, we highlight below several avenues for future research.

\subsubsection{Language Extensibility}\label{sec:language-extensibility}

\ivy{} is designed to be extensible:
support for each specification language is defined through a standardized interface, which includes a JSON Schema describing the syntax, a JavaScript rendering function for the language, and rewrite rules to help users abstract specifications. Our implementation currently supports a small set of languages (Vega, Vega-Lite, Atom\footnoteWithIndent{While implementing support for Atom we extracted the language in the original application into a standalone library: \url{https://www.npmjs.com/package/unit-vis}
}, and a simple table language), which, in future work, we would like to increase so as to support a greater variety of tasks.

Our grammar-agnostic template framework provides a standard set of abstraction and data manipulation mechanisms---which may reduce the need for grammar designers to define their own---and novel UI features for exploring candidate templates (catalog search) and encodings (fan outs)---which may facilitate more efficient and consistent exploration.
Our system, furthermore, hoists the burden of data transformation out of the rendering grammar (albeit with a currently-limited set of transforms), which would otherwise require each grammar to implement its own data manipulation logic.
As users move between templates (specified in possibly different grammars), their settings (including filters) are mapped from one template to the next via role and order-based heuristics.
Future work could
enable translation between supported grammars,
which could yield opportunities for education and portability.

Despite the benefits of language-independent functionality, there are also benefits to taking domain-specific knowledge into account.
Language-specific rewrite rules---part of the extension interface, described above---are one such example.
Language-specific knowledge could further be used for recommendation, as well as data manipulation and presentation concerns.
For instance, when a data field is dragged to a drop zone in Lyra~\cite{carr_lyra_2014} the appropriate type of scale is automatically inferred~\cite{satyanarayan_critical_2020}.
Lyra is able to offer this functionality because it has a model of the grammar being manipulated, a functionality which our approach currently lacks.

A lack of context and content-aware automated guidance is a key limitation of our design.
Yet, it should be possible to identify a richer extensibility API, while still allowing each language to benefit from the abstraction and UI concerns shared by all.
Such an API would enable us to combine domain-specific chart recommendation (\secref{sec:recommendation})
with \ivy{}'s domain-independent type-based exploration (\secref{sec:emergent}), as well as embrace new interaction modalities.

\subsubsection{Validation in Visual Analytics}

An important question for system designers is how to help users conduct safe visual analysis~\cite{zhao_safe_2017}.
Analysts can deceive themselves with statistical traps~\cite{pu_garden_2018}, visualization hallucinations~\cite{kindlmann_algebraic_2014}, or false graphical inferences~\cite{ wickham_graphical_2010}.

We believe that our type-based template search will dovetail with a metamorphic testing~\cite{mcnutt_surfacing_2020} based validation approach: by varying the parameters of a template and comparing the resulting images, a validation system could automatically identify errors at the intersection of data and encoding.
Similarly, we suggest that templates likely offer an opportune medium for applying visualization linting~\cite{mcnutt_linting_2018, hopkins_visualint_2020} to a visual analytics context, as the types expose the specific arguments over which analysis could be conducted.
Furthermore, the fan out interaction could be extended to allow not only juxtaposition of variants, but also their layering~\cite{lunzer2014livelyr}, enabling visual sanity checks to the robustness of parameter selections~\cite{correll2018looks}.

\subsubsection{Integrated Visualization Editors}

A primary focus in this paper has been supporting data exploration tasks, but there are a constellation of other tasks that users perform in visualization tools.
Users must carry out tasks ``before'' exploration (such as model building and data cleaning) and ``after'' exploration (such as annotation and presentation).
Although the designs of many \va{} systems assume that data has been cleaned and processed prior to analysis~\cite{satyanarayan_critical_2020}, in practice these tasks are often interleaved and iterated.

Integrating these tasks within a unified system may thus reap potential benefits.
Data manipulation tasks, for instance, might be better facilitated by the incorporation of ideas from dataflow programming and spreadsheets,
as well as \emph{programming-by-example} techniques to help with data wrangling~\citep{kandel_wrangler_2011,foofah, evensen2020ruler}.
To more fully support presentation tasks, it would be fruitful to extend our combination of modalities to include \emph{visual builders}~\cite{grammel_survey_2013}---which offer a variety of direct manipulation features~\cite{shneiderman_direct_1983, saket_investigating_2019} for creating charts---and \emph{visualization by demonstration}~\cite{saket_liger_2019, saket_visualization_2016, wang_visualization_2019, zong2021lyra}.

For each of these tasks, developing rich graphical interactions---while maintaining a ``bidirectional'' connection to the underlying textual representation---is an exciting research challenge: to integrate what is typically a vast divide between text and GUI based analytics systems.
The approach pursued in this paper provides a step in this longer-term direction.

\section{Acknowledgments}
We thank our anonymous reviewers for their helpful comments, as well as our study participants for their participation and invaluable insights. We also thank
Will Brackenbury,
Michael Correll,
Galen Harrison,
Brian Hempel,
Gordon Kindlmann,
and Katy Koenig
for their commentary and support.

\bibliographystyle{ACM-Reference-Format}
\bibliography{main}

\newpage

\appendix

\begin{figure*}[!t]
      \begin{minipage}{\doubleColFigWidth}
            \centering
            \includegraphics[width=0.9\linewidth]{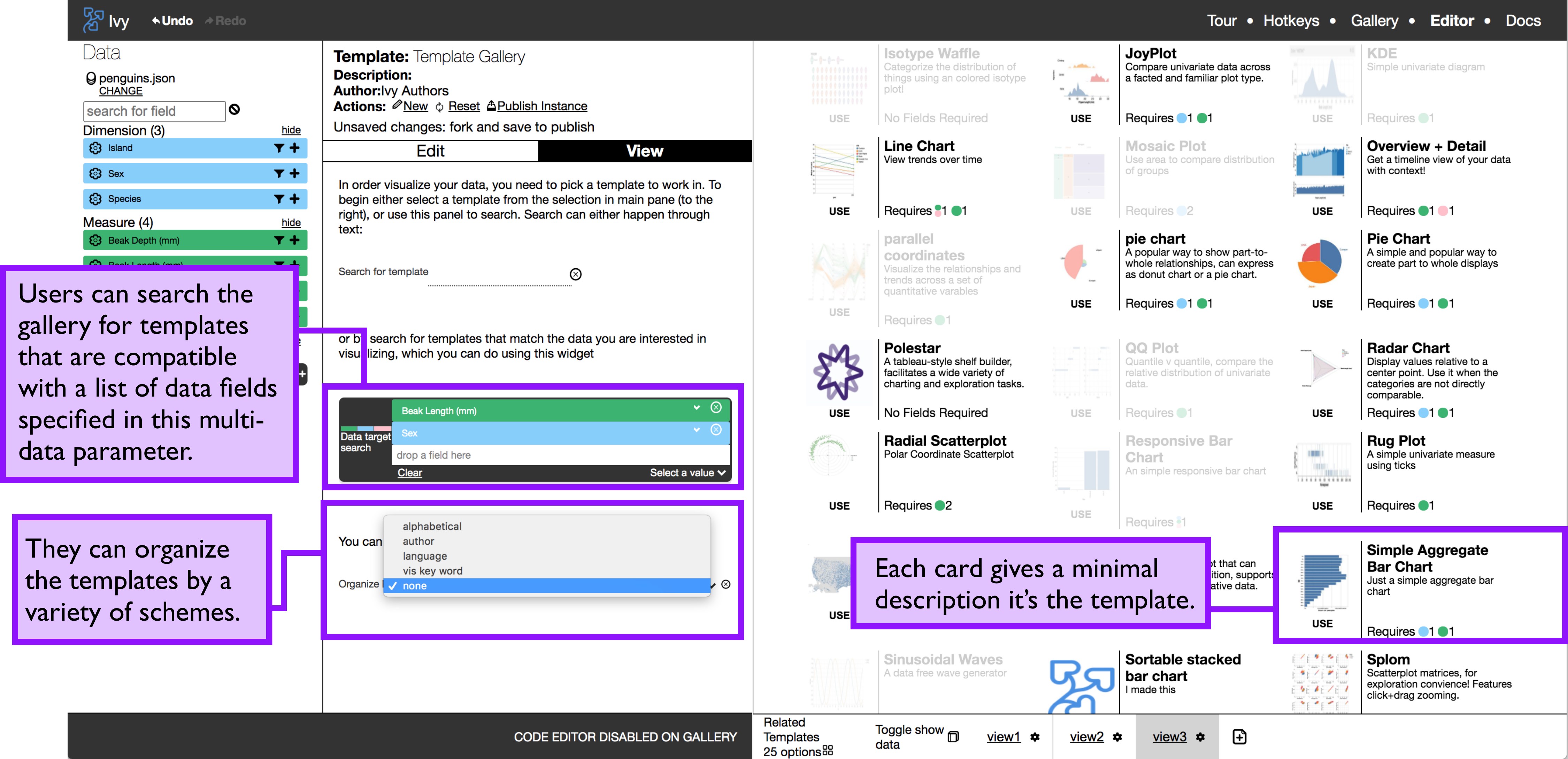}
            \caption{
                  Users can search for templates via text or catalog search in the gallery, which is a set of user created templates (hosted on a communal server) and system templates.
            }
            \Description[A view of the gallery.]{A view of the gallery. The gallery is annotated highlighting various features of the gallery. }
            \label{fig:gallery}
      \end{minipage}
\end{figure*}

\begin{figure*}[!t]
    \begin{minipage}{\doubleColFigWidth}
        \centering
        \includegraphics[width=0.9\linewidth]{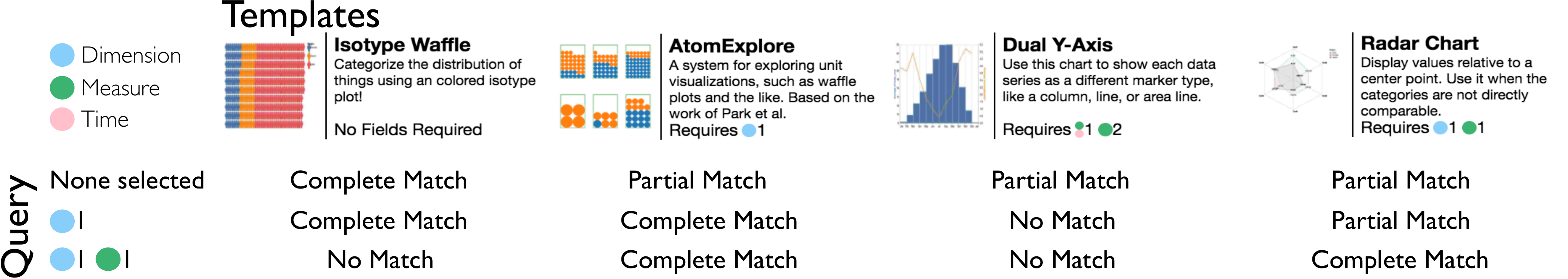}
        \caption{
            Catalog search (\secref{sec:ui-catalog-search}) compares templates to a set of data fields, which can yield a partial match (no conflict but unrenderable), a complete match (renderable), or no match (conflict).
        }
        \Description[The tabular description of the matching algorithm]{The tabular description of the matching algorithm. The tabular description compares four templates (horizontal) against three different role-based queries, noting them as Complete Matches/Partial Matches/No matches.}
        \label{fig:catalog-search-explain}
    \end{minipage}
\end{figure*}

\section{Appendix}

In this appendix we expand upon several elements of the design, implementation, and evaluation of \ivy{} discussed in the main paper.

\subsection{Catalog Search Matching Heuristic}\label{sec:matching}

\figref{fig:gallery} shows the gallery of \ivy{} templates in our current implementation.
Here we provide a more precise description of the heuristic used to perform the template matching in our catalog search.

For a template $t$ (among the set of templates $T$) with a set of data parameters $d$, (where each parameter has a set of allowed types $d_i\tau$), and a search $S$ consisting of a set of data columns $\{c_i\}$ which each have a single type $c_i\tau$, then $S$ is a \textit{partial match} for $t$ if there is an injective mapping $m$ between them such that
\begin{equation}
      (\exists m)(\forall c_i \in S)(\exists d_j \in d)(m: c_i \rightarrow d_j: c_i\tau\in d_j\tau)
\end{equation}
Correspondingly $S$ is a \textit{full match} for $t$ if
\begin{equation}
      (\exists m)(\forall rd_i \in rd)(\exists c_j \in S)(m: c_i \rightarrow d_j: c_i\tau\in rd_j\tau)
\end{equation}
where we define the required set of data param in $t$ as $rd\subseteq d$. This check is bounded by the size of $S $ and $T$, so a search across the templates will take  $\mathcal{O}(|S||T|)$.
We illustrate this  in \figref{fig:catalog-search-explain}.

\subsection{Implementation Details}

\system{} is a TypeScript React-Redux application.  We were motivated to use React as the basis of our application as it provides an opinionated approach on how to build additional renderers, which is important for our approach to extensibility.
Our template server is a cloud-based node.js server backed by PostgreSQL.

As described in \secref{sec:language-extensibility} our system is designed to be extensible:
support for each specification language is defined through an extension interface comprising metadata (such as a JSON Schema describing the syntax), a React component~\citeAppendix{facebook_react_nodate} that exposes the rendering function of the language, and rewrite rule definitions that help users abstract specifications into templates.
While Ivy currently supports a relatively limited class of rewrite rules, future work could extend the approach with more expressive languages for transforming structured data (such as in \toolName{CDuce}~\citeAppendix{benzaken_cduce_2003} and \toolName{XDuce}~\citeAppendix{hosoya_xduce_2003}).

Our current implementation supports four languages---Vega, Vega-Lite, Atom,
and a toy data table language---which serves as
a limited demonstration of the validity of our extensible approach.
As JSON-mediated visualization grammars continue to gain popularity, additional languages will inevitably emerge to solve problems unaddressed in prior efforts.
Future languages could support more complex rendering schemes, focusing on particular domains such as geospatial analytics~\citeAppendix{he_keplergl_nodate}, 3D visual analytics, pivot tables (perhaps simplifying the language of VizQL~\cite{hanrahan_vizql_2006}),  or even on the chart recommendation language CompassQL~\citeAppendix{wongsuphasawat_towards_2016}---which would enable task-specific variations of Voyager~\cite{wongsuphasawat_voyager_2017}.

\begin{figure}[b]
    \begin{minipage}{\oneColFigWidth}
        \centering
        \includegraphics[width=\linewidth]{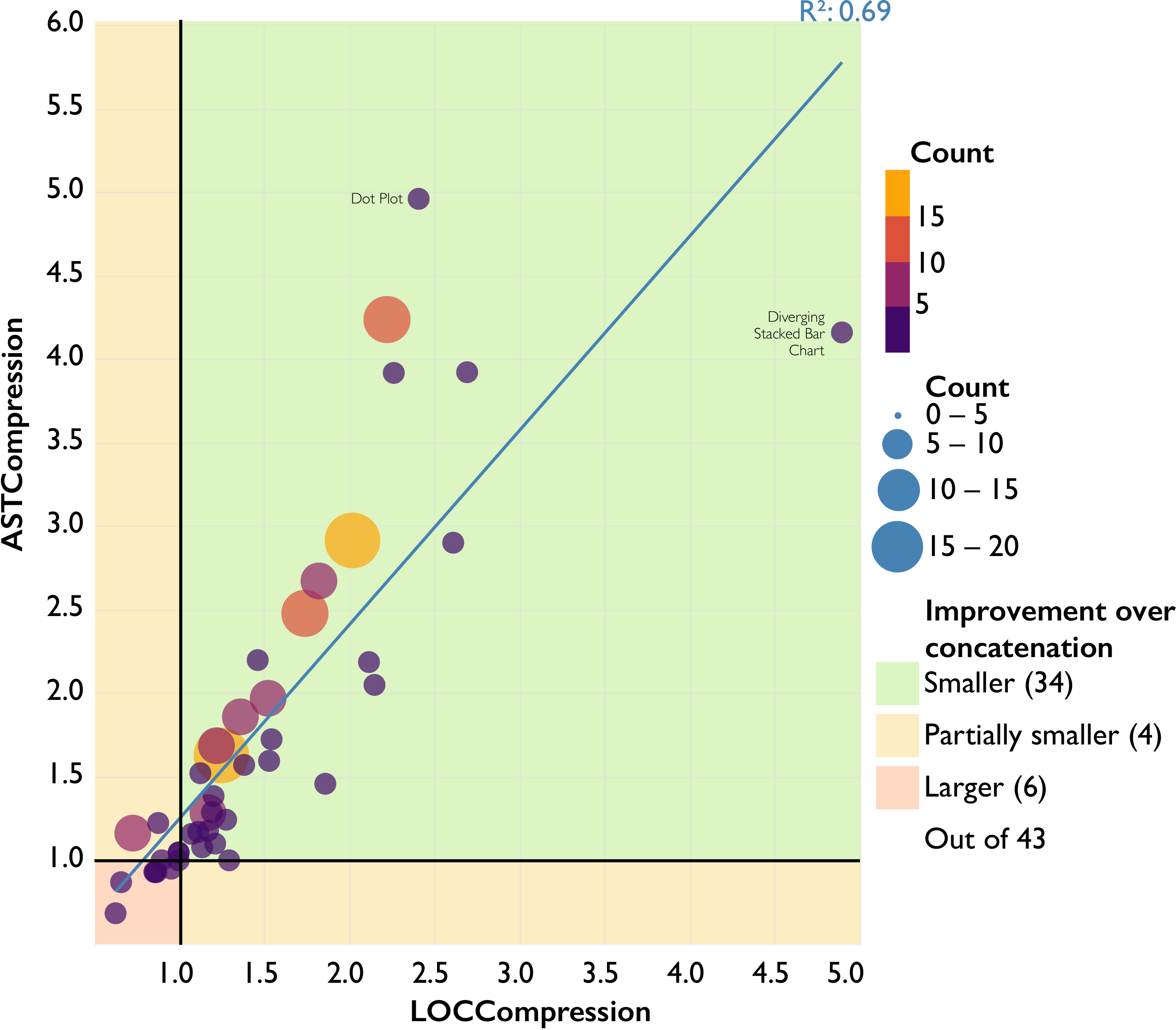}
        \caption{
            The concatenation ratios across our reproduction of the Vega-Lite gallery between the compression of the Abstract Syntax Tree (a stand-in for complexity minimization) and Line of Code Compression (a stand-in simplification). Higher is better in both cases.
        }
        \label{fig:compression-fig}
        \Description[]{A scatterplot of the compression ratios for the vega-lite gallery reproduction. LOC Compression is on the x-axis and AST Compression on the y-axis}
    \end{minipage}
\end{figure}

\subsection{Templates for Vega-Lite Gallery}

As described in \secref{sec:vega-lite}, we aimed to factor the Vega-Lite examples into templates in reasonable ways, a heuristic which was guided by the minimization of complexity and the maximization of simplification. We illustrate these metrics in \figref{fig:compression-fig}.
The metrics in this figure are computed following \autoref{eqn:concatRatio}.
Most templates exhibit a compression greater than 1, indicating that the template is better than simply concatenating the examples together. Those that do worse tend to have particular affordances to the template usable and also tend to only cover a single example, limiting their compression. See \osf{} for further details.

\subsection{Templates to Reproduce Chart Choosers}

\begin{figure}[b]
    \begin{minipage}{\oneColFigWidth}
        \centering
        \includegraphics[width=\linewidth]{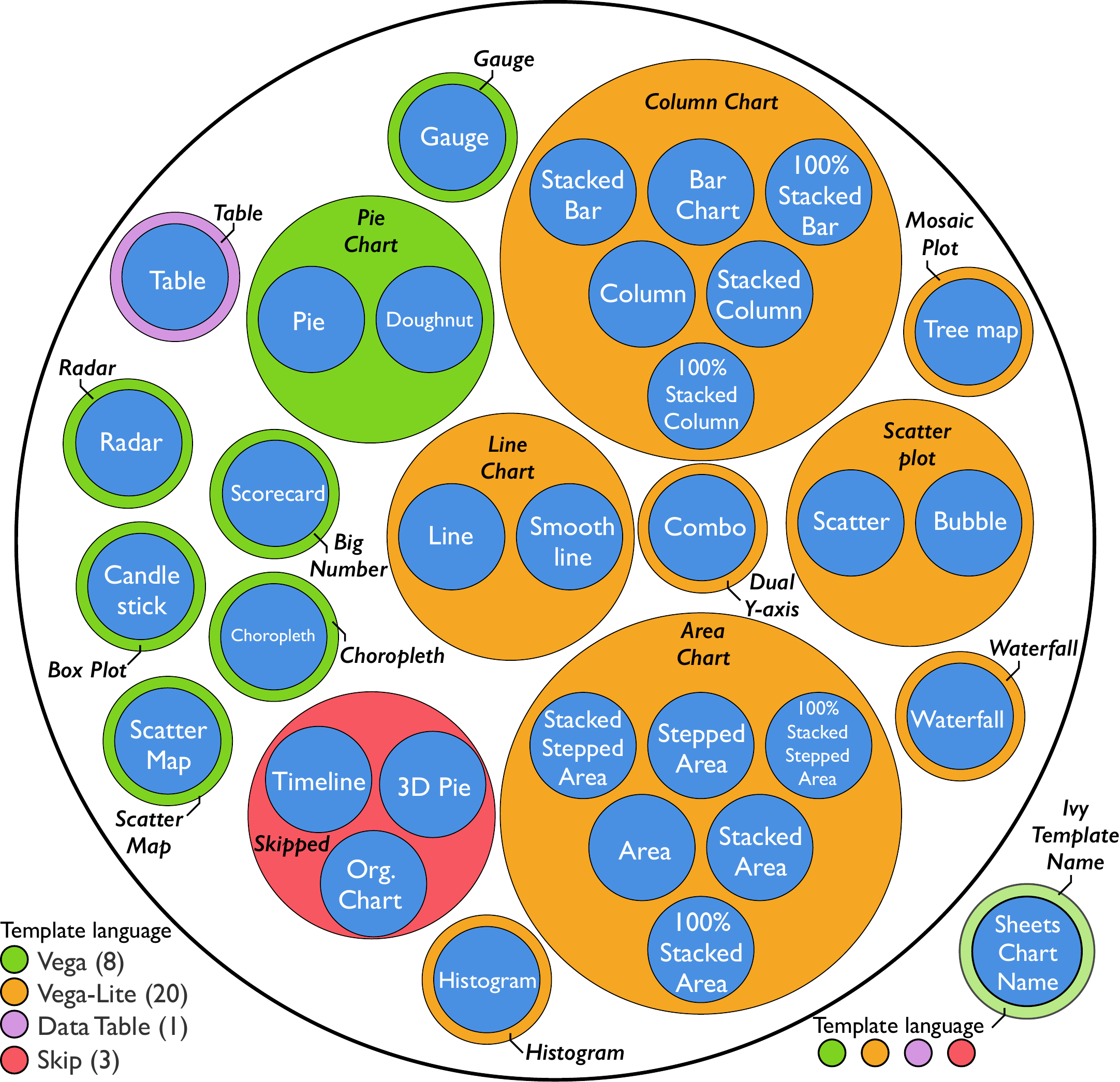}
        \caption{
            We created 16 \system{} templates that reconstruct the functionality of 29 of the 32 charts in the Google Sheets chart chooser, a $1.8$x compression.
        }
        \label{fig:expressiveness}
        \Description[An annotated circepack diagram]{An annotated circepack diagram showing which chart chooser options were captured by which ivy templates. A legend on the lower left shows the numerical breakdown of these templates/chart chooser matchings.}
    \end{minipage}
\end{figure}

We describe here in greater detail our templatization of the Google Sheets chart chooser (described in \secref{sec:google-sheets}), as well as an additional chart corpus provided by Russell \citeAppendix{russell_simple_2016}.
The latter fell outside the narrative in the main body of the paper, but we include it here as an example of the chart making culture in one particular organization.
\tabref{table:appendix-matching} summarizes the resulting templates described below.

\subsubsection{Google Sheets}

The Sheets chart chooser consists of 32 options.
We reproduce 29 of these through 16 templates, as summarized in \figref{fig:expressiveness}.
We skipped ``3D Pie'' charts because there is not yet a dominant grammar for browser-based non-VR 3D visual analytics, although several recent works have put forward interesting approaches \citeAppendix{sicat_dxr_2018, butcher_vria-framework_2019}.
We skipped ``Org. Charts'' because they fall outside of our tabular data model, requiring a hierarchical one.
Finally, we skipped ``Timelines'' because we do not currently support the data manipulations required to support textual annotations as required by this chart form.
Each of these deficiencies could be addressed in future work, such as by extending the range of languages supported.

\subsubsection{Russell Survey}

This survey\citeAppendix{russell_simple_2016} of internal presentations at Google included approximately 1,300 charts, which were grouped into 15 distinct visual forms.
We reproduce 10 of these through 11 templates.
There are more templates than charts because, following Sheets, we split Russell's ``Map'' into two templates, ``Country Choropleth'' and ``Scatter Map''.
Of the 5 charts from this survey we skipped, 3 involve a non-tabular data model, 1 requires domain-specific data (Lam \etals~SessionViewer~\citeAppendix{lam_session_2007}), and 1 uses annotations.
Among the 16 plus 11 templates described, 18 are distinct.

While informative, this selection covers one particular analytic culture and one family of tool's designs. For instance, Russell's review found SessionViewer~\citeAppendix{lam_session_2007}, a system for understanding web search usage behaviors, made up 2.1\% of the review corpus. A review of a different corpus would likely yield a different selection of charts.
Furthermore, this selection of charts is also a symptom of software availability. To wit: unit visualizations tend to be uncommon because few systems tend to support them \cite{park_atom_2018}.

\begin{table*}[ht]
    \caption{We now give additional detail regarding the overlap between our coverage of the Google Sheets Chart Chooser Options and the charts described in Russell's survey. Options that are unsupported under the current set of grammars are marked as \colorbox{red!35!}{Skip}. Each template was created by starting from a blank template, from using Polestar to approximate the right behavior, or by abstracting an example found online.
    }
    \begin{tabular}{lllccll} \toprule
        Russel Chart               & Sheets Chart                     & Ivy Template             & Russell                          & Sheets                           & Language                   & Creation Method            \\ \toprule
        line graph                 & Smooth Line Chart                & Line Chart               & \colorbox{green!15!}{\checkmark} & \colorbox{green!15!}{\checkmark} & Vega-lite                  & Polestar
        \\
        line graph                 & Line Chart                       & Line Chart               & \colorbox{green!15!}{\checkmark} & \colorbox{green!15!}{\checkmark} & Vega-lite                  & Polestar
        \\
        histogram                  & Histogram chart                  & Histogram                & \colorbox{green!15!}{\checkmark} & \colorbox{green!15!}{\checkmark} & Vega-lite                  & Polestar
        \\
        table                      & Table chart                      & Table                    & \colorbox{green!15!}{\checkmark} & \colorbox{green!15!}{\checkmark} & Data table                 & Blank
        \\
        pie                        & Pie chart                        & Pie chart                & \colorbox{green!15!}{\checkmark} & \colorbox{green!15!}{\checkmark} & Vega                       & Example
        \\
        pie                        & Doughnut chart                   & Pie chart                & \colorbox{green!15!}{\checkmark} & \colorbox{green!15!}{\checkmark} & Vega                       & Example
        \\
        stack histogram            & Column chart                     & Column Chart             & \colorbox{green!15!}{\checkmark} & \colorbox{green!15!}{\checkmark} & Vega-lite                  & Polestar
        \\
        stack histogram            & Stacked column chart             & Column Chart             & \colorbox{green!15!}{\checkmark} & \colorbox{green!15!}{\checkmark} & Vega-lite                  & Polestar
        \\
        stack histogram            & 100\% stacked column chart       & Column Chart             & \colorbox{green!15!}{\checkmark} & \colorbox{green!15!}{\checkmark} & Vega-lite                  & Polestar
        \\
        stack histogram            & Bar chart                        & Column Chart             & \colorbox{green!15!}{\checkmark} & \colorbox{green!15!}{\checkmark} & Vega-lite                  & Polestar
        \\
        stack histogram            & Stacked bar chart                & Column Chart             & \colorbox{green!15!}{\checkmark} & \colorbox{green!15!}{\checkmark} & Vega-lite                  & Polestar
        \\
        stack histogram            & 100\% stacked bar chart          & Column Chart             & \colorbox{green!15!}{\checkmark} & \colorbox{green!15!}{\checkmark} & Vega-lite                  & Polestar
        \\
        box plot                   & Candlestick chart                & Candle stick             & \colorbox{green!15!}{\checkmark} & \colorbox{green!15!}{\checkmark} & Vega                       & Example
        \\
        scatterplot                & Scatter chart                    & Scatterplot              & \colorbox{green!15!}{\checkmark} & \colorbox{green!15!}{\checkmark} & Vega-lite                  & Blank
        \\
        scatterplot                & Bubble chart                     & Scatterplot              & \colorbox{green!15!}{\checkmark} & \colorbox{green!15!}{\checkmark} & Vega-lite                  & Blank
        \\
        map                        & Geo chart                        & Country Choropleth       & \colorbox{green!15!}{\checkmark} & \colorbox{green!15!}{\checkmark} & Vega                       & Example
        \\
        map                        & Geo chart with markers           & Scatter Map              & \colorbox{green!15!}{\checkmark} & \colorbox{green!15!}{\checkmark} & Vega                       & Blank
        \\
        timeline                   & Timeline chart                   & \colorbox{red!35!}{Skip} & \colorbox{green!15!}{\checkmark} & \colorbox{green!15!}{\checkmark} & \colorbox{orange!50!}{N/A} & \colorbox{orange!50!}{N/A}
        \\
        pie                        & 3D pie chart                     & \colorbox{red!35!}{Skip} & \colorbox{green!15!}{\checkmark} & \colorbox{green!15!}{\checkmark} & \colorbox{orange!50!}{N/A} & \colorbox{orange!50!}{N/A}
        \\
        heatmap                    & \colorbox{orange!50!}{N/A}       & Heatmap                  & \colorbox{green!15!}{\checkmark} & \colorbox{red!35!}{$\times$}     & Vega-lite                  & Example
        \\
        sunburst                   & \colorbox{orange!50!}{N/A}       & Sunburst                 & \colorbox{green!15!}{\checkmark} & \colorbox{red!35!}{$\times$}     & Vega                       & Example
        \\
        arc/node graph             & \colorbox{orange!50!}{N/A}       & \colorbox{red!35!}{Skip} & \colorbox{green!15!}{\checkmark} & \colorbox{red!35!}{$\times$}     & \colorbox{orange!50!}{N/A} & \colorbox{orange!50!}{N/A}
        \\
        SessionView                & \colorbox{orange!50!}{N/A}       & \colorbox{red!35!}{Skip} & \colorbox{green!15!}{\checkmark} & \colorbox{red!35!}{$\times$}     & \colorbox{orange!50!}{N/A} & \colorbox{orange!50!}{N/A}
        \\
        Sankey                     & \colorbox{orange!50!}{N/A}       & \colorbox{red!35!}{Skip} & \colorbox{green!15!}{\checkmark} & \colorbox{red!35!}{$\times$}     & \colorbox{orange!50!}{N/A} & \colorbox{orange!50!}{N/A}
        \\
        force vector               & \colorbox{orange!50!}{N/A}       & \colorbox{red!35!}{Skip} & \colorbox{green!15!}{\checkmark} & \colorbox{red!35!}{$\times$}     & \colorbox{orange!50!}{N/A} & \colorbox{orange!50!}{N/A}
        \\
        \colorbox{orange!50!}{N/A} & Organizational chart             & \colorbox{red!35!}{Skip} & \colorbox{red!35!}{$\times$}     & \colorbox{green!15!}{\checkmark} & \colorbox{orange!50!}{N/A} & \colorbox{orange!50!}{N/A}
        \\
        \colorbox{orange!50!}{N/A} & Combo chart                      & Dual Y-axis              & \colorbox{red!35!}{$\times$}     & \colorbox{green!15!}{\checkmark} & Vega-lite                  & Example
        \\
        \colorbox{orange!50!}{N/A} & Waterfall chart                  & Waterfall                & \colorbox{red!35!}{$\times$}     & \colorbox{green!15!}{\checkmark} & Vega-lite                  & Example
        \\
        \colorbox{orange!50!}{N/A} & Radar chart                      & Radar                    & \colorbox{red!35!}{$\times$}     & \colorbox{green!15!}{\checkmark} & Vega                       & Example
        \\
        \colorbox{orange!50!}{N/A} & Gauge chart                      & Gauge                    & \colorbox{red!35!}{$\times$}     & \colorbox{green!15!}{\checkmark} & Vega                       & Example
        \\
        \colorbox{orange!50!}{N/A} & Scorecard chart                  & BigNumber                & \colorbox{red!35!}{$\times$}     & \colorbox{green!15!}{\checkmark} & Vega                       & Blank
        \\
        \colorbox{orange!50!}{N/A} & Tree map chart                   & Mosaic Plot              & \colorbox{red!35!}{$\times$}     & \colorbox{green!15!}{\checkmark} & Vega-lite                  & Example
        \\
        \colorbox{orange!50!}{N/A} & Area chart                       & Area chart               & \colorbox{red!35!}{$\times$}     & \colorbox{green!15!}{\checkmark} & Vega-lite                  & Polestar
        \\
        \colorbox{orange!50!}{N/A} & Stacked area chart               & Area chart               & \colorbox{red!35!}{$\times$}     & \colorbox{green!15!}{\checkmark} & Vega-lite                  & Polestar
        \\
        \colorbox{orange!50!}{N/A} & 100\% stacked area chart         & Area chart               & \colorbox{red!35!}{$\times$}     & \colorbox{green!15!}{\checkmark} & Vega-lite                  & Polestar
        \\
        \colorbox{orange!50!}{N/A} & Stepped area chart               & Area chart               & \colorbox{red!35!}{$\times$}     & \colorbox{green!15!}{\checkmark} & Vega-lite                  & Polestar
        \\
        \colorbox{orange!50!}{N/A} & Stacked stepped area chart       & Area chart               & \colorbox{red!35!}{$\times$}     & \colorbox{green!15!}{\checkmark} & Vega-lite                  & Polestar
        \\
        \colorbox{orange!50!}{N/A} & 100\% stacked stepped area chart & Area chart               & \colorbox{red!35!}{$\times$}     & \colorbox{green!15!}{\checkmark} & Vega-lite                  & Polestar
    \end{tabular}
    \label{table:appendix-matching}
\end{table*}

\ifonecol
\else
    \newpage
    ~
    \newpage
    ~
    \newpage
\fi

\subsection{User Study Prompts}

Here we provide the text of the tasks involved in the user study. The full study instrument can be found in the supplementary materials. These questions were divided into two sections, \emph{Tutorial}, which involved substantial guidance, and \emph{Independent}, which were more freeform. These tasks were selected because they were similar to tasks that one might address in similar systems.

\subsubsection{Tutorial Tasks}

\begin{enumerate}[leftmargin=*]
      \item  In this task you will make a small multiple log-log scatter plot of happiness vs population for 2015 colored and faceted by region (such as by row or column) where tooltipping shows (among other data) the name of the country. To do so you make use of the Polestar template. Start by finding the Polestar template in the gallery, navigate to it. Now fill in the appropriate fields for X and Y. To make the tooltip reveal useful information, place the Country field onto the detail target. Set the Region to Column as well. Don't forget to filter the appropriate year.
            What correlation can you see? Give your answer in plain text.

      \item  In this task we will make a SPLOM (scatter plot matrix) for our dataset. Start by creating a new view, and navigate to the Polestar template. Place the row and column cards on the x and y data targets respectively. Next, select 3 measures of interest (you decide!), place each of them in both of the ``row'' and ``column'' multi targets which are under the ``meta columns section''. Now click the lighting bolt next the Color field and select 3 dimensions of interest (you decide!). Just as before it might be helpful to place Country into Detail. What correlations can you find? Give your answer in plain text.

      \item  Next you will make use of a particular template from the gallery. Specifically, you will make a radial scatterplot. This task is a little different in that you will use a different dataset. Open the gallery in a new tab and find the template that will allow you to make a radial scatterplot. Once there select the penguins dataset. What is the most interesting combination of variables you can find? Can you use fanout to effectively move through these options? These plots are a little big. Why don't you try to navigate to the code body and change their height and width to be something a little more reasonable? Copy the code from the output into the below box.

      \item  Next you will try out making a template, specifically let's make a heatmap. Load up the happiness dataset once again. Once again we will start with Polestar. Start by placing the CountryType and GovernmentType variables on to the X and Y targets. Then place the happiness field onto the Color Field. Feel free to adjust the heatmap as you like. You should probably change the mark type to make it more heatmap like! When you are ready, click the ``Fork'' button and select ``Just output''. Click the suggestion for CountryType that creates and configures a new field, this will create a new datatarget for this field while keeping the current graphic in place. Select the gear to the right of this new field and give it an informative field name, and select only the appropriate field data type. Do the same for GovernmentType. Next, let's make our heatmap be better able to describe various aggregates. Create a new List widget and add options for each of the aggregation types (the ones that make the most sense are probably count and distinct, but you are welcome to use others, see \url{https://vega.github.io/vega-lite/docs/}). Give the widget a descriptive name (no spaces though!) and replace the word ``count'' in the Body with your new name wrapped in brackets, like so ``[YOUR NEW NAME]''. You should now be able to switch through various aggregations (or fan across them). Finally, give your template a name and description, and then publish it!
\end{enumerate}

\subsubsection{Independent Tasks}
\begin{enumerate}[leftmargin=*]
      \item  How many rows are in this dataset? How many countries are represented in this dataset? How many countries are in each region? Please give you answers in plain text. (You can answer the last question using the JSON output.)

      \item  Create a new template by adapting \url{https://vega.github.io/vega-lite/examples/boxplot_minmax_2D_vertical.html} to this dataset in a useful manner. The choice of columns is up to you. Try adding height and width sliders. Try enabling switching between types of box plot e.g. \url{https://vega.github.io/vega-lite/examples/boxplot_2D_vertical.html)}. Copy the body of your template into the space below.

      \item  Please make charts that answer the following questions. What is the global trend in corruption? Do bigger countries tend to be happier? Use code from the output for your answers.

      \item  Please combine the following examples drawn from the Vega-Lite gallery into an Ivy template: 1. \url{https://vega.github.io/vega-lite/examples/bar_aggregate.html} 2.\url{https://vega.github.io/vega-lite/examples/bar_aggregate_sort_by_encoding.html}. The particular choice of columns and features is up to you. Hint: the big difference is that one is sorted and the other is not! If you are having trouble with this task it may be helpful to check out the documentation of the template language found on the web page. Answer this question by giving the body of the template you've created.
\end{enumerate}


\bibliographystyleAppendix{ACM-Reference-Format}
\bibliographyAppendix{main}

\end{document}